\documentclass[%
nofootinbib,
preprint,
superscriptaddress,
 amsmath,amssymb,
prc,
floatfix,
numbers,https://www.overleaf.com/project/612ed19b5e8a3b350599e950
]{revtex4-2}

\RequirePackage[T1]{fontenc}

\RequirePackage{graphicx}

\graphicspath{{figures/}} 

\usepackage{epstopdf}

\usepackage{amsmath}
\usepackage{amsfonts}
\usepackage{amssymb}
\usepackage{textcomp}
\usepackage[utf8]{inputenc}
\usepackage{graphicx}

\usepackage[margin=2.5cm]{geometry}

\usepackage{slashed}

\usepackage{feynmp-auto}

\begin{document}

\title{Jefferson Lab Hall C: Precision Physics at the Luminosity Frontier}

\author{J. Benesch       } 
\affiliation{Thomas Jefferson National Accelerator Facility, Newport News, VA 23606, USA} 
\author{V.~Berdnikov          } 
\affiliation{The Catholic University of America, Department of Physics, Washington, DC, 20064, USA} 
\author{P. Brindza        } 
\affiliation{Thomas Jefferson National Accelerator Facility, Newport News, VA 23606, USA}
\author{S.~Covrig~Dusa } 
\affiliation{Thomas Jefferson National Accelerator Facility, Newport News, VA 23606, USA}
\author{D.~Dutta}
\affiliation{Mississippi State University, Mississippi State, MS 39762, USA}
\author{D.~Gaskell}
\affiliation{Thomas Jefferson National Accelerator Facility, Newport News, VA 23606, USA}
\author{T.~Gogami}
\affiliation{Department of Physics, Graduate School of Science, Kyoto University, Kyoto, Kyoto 606-8502, Japan}
\author{J.M.~Grames}
\affiliation{Thomas Jefferson National Accelerator Facility, Newport News, VA 23606, USA}
\author{D.J.~Hamilton }
\affiliation{University of Glasgow, Glasgow, G12 8QQ, UK}
\author{D.W.~Higinbotham }
\affiliation{Thomas Jefferson National Accelerator Facility, Newport News, VA 23606, USA}
\author{T.~Horn          } 
\affiliation{The Catholic University of America, Department of Physics, Washington, DC, 20064, USA}
\affiliation{Thomas Jefferson National Accelerator Facility, Newport News, VA 23606, USA} 
\author{G.M.~Huber       } 
\affiliation{University of Regina, Regina, SK  S4S~0A2, Canada}
\author{M.~K.~Jones     }
 \affiliation{Thomas Jefferson National Accelerator Facility, Newport News, VA 23606, USA}
\author{C.~Keith      }
 \affiliation{Thomas Jefferson National Accelerator Facility, Newport News, VA 23606, USA}
\author{C.~Keppel       }
 \affiliation{Thomas Jefferson National Accelerator Facility, Newport News, VA 23606, USA}
\author{E.R.~Kinney      }
\affiliation{University of Colorado, Boulder, CO 80309-0390, USA}
\author{W.B.~Li          }
\affiliation{William \& Mary,  VA 23185, USA}
\affiliation{Center for Frontiers in Nuclear Science, Stony Brook University, Stony Brook, NY 11794, USA}
\affiliation{Department of Physics and Astronomy, Stony Brook University, Stony Brook, NY 11794, USA}
\affiliation{Thomas Jefferson National Accelerator Facility, Newport News, VA 23606, USA}
\author{Shujie~Li}
\affiliation{LBNL, Berkeley, CA 94720, USA}
\author{N. Liyanage        }
\affiliation{University of Virginia, Department of Physics, Charlottesville, VA 22904, USA}
\author{E.~Long          } 
\affiliation{University of New Hampshire, Department of Physics and Astronomy, Durham, NH 03824, USA}
\author{D.J.~Mack} 
\email{Corresponding Author E-mail: mack@jlab.org}
\affiliation{Thomas Jefferson National Accelerator Facility, Newport News, VA 23606, USA}
\author{B.~Metzger} 
\affiliation{Thomas Jefferson National Accelerator Facility, Newport News, VA 23606, USA}
\author{C.~Mu\~noz~Camacho}
\affiliation{Université Paris-Saclay, CNRS, IJCLab, 91406 Orsay, France}
\author{S.N.~Nakamura} 
\affiliation{Department of Physics, Graduate School of Science, Tohoku University, Sendai, Miyagi 980-8578, Japan}
\author{B.~Sawatzky} 
\affiliation{Thomas Jefferson National Accelerator Facility, Newport News, VA 23606, USA}
\author{K.~Slifer} 
\affiliation{University of New Hampshire, Department of Physics and Astronomy, Durham, NH 03824, USA}
\author{H.~Szumila-Vance}
\affiliation{Thomas Jefferson National Accelerator Facility, Newport News, VA 23606, USA}
\author{A.S.~Tadepalli} 
\affiliation{Thomas Jefferson National Accelerator Facility, Newport News, VA 23606, USA}
\author{L.~Tang} 
\affiliation{Thomas Jefferson National Accelerator Facility, Newport News, VA 23606, USA}
\affiliation{Hampton University, Hampton, VA 23669, USA}
\author{B.~Wojtsekhowski} 
\affiliation{Thomas Jefferson National Accelerator Facility, Newport News, VA 23606, USA}
\author{S.A.~Wood} 
\affiliation{Thomas Jefferson National Accelerator Facility, Newport News, VA 23606, USA}
\affiliation{Old Dominion University, Norfolk, VA 23529, USA}

\date{\today}


\begin{abstract}

Over the last three decades, Hall C has been a key contributor to progress 
in the understanding of hadron structure and interactions. 
An outline of a potential future Hall C 
physics program focused on precision measurements of small cross sections is presented.
A detailed overview of this unique facility,
whose flexible configuration allows many opportunities for new experimental equipment that help address a wide range of questions in hadronic physics,
is included as well.  

\end{abstract}

\maketitle
 
%


\pagebreak

\tableofcontents

\pagebreak

\section{Introduction}


Teasing out exactly how hadrons emerge from quarks and gluons carrying color charge is a major focus area of nuclear science. This
key question is being addressed at Jefferson Lab (JLab) with a coordinated and systematic mapping of
the internal landscape of hadrons.
The importance of the hadron physics experiments in the energy domain of JLab covering center of mass energies between 3.5 and nearly 5 GeV was presented in a series of 
documents which supported the upgrade of the Jefferson Lab accelerator from  6~GeV to 12~GeV \cite{PAC18, WP1, WP2}. A recent paper \cite{WP3} summarizes the scientific accomplishments in all 4 Halls at Jefferson Lab that have been completed thus far with the 12~GeV upgrade. In this context, Hall C has been providing the highest precision measurements of the separated electromagnetic responses critical for testing theories of hadronic forces. 
In conjunction with accurate beam diagnostics, the Hall C focusing spectrometers' high momentum reach, rigid connection to a sturdy pivot, and reproducible magnetic properties have allowed control of systematic uncertainties at the 1\% level, 
enabling these delicate separations. 

\begin{figure*}
\centering
\includegraphics[width=0.95\textwidth]{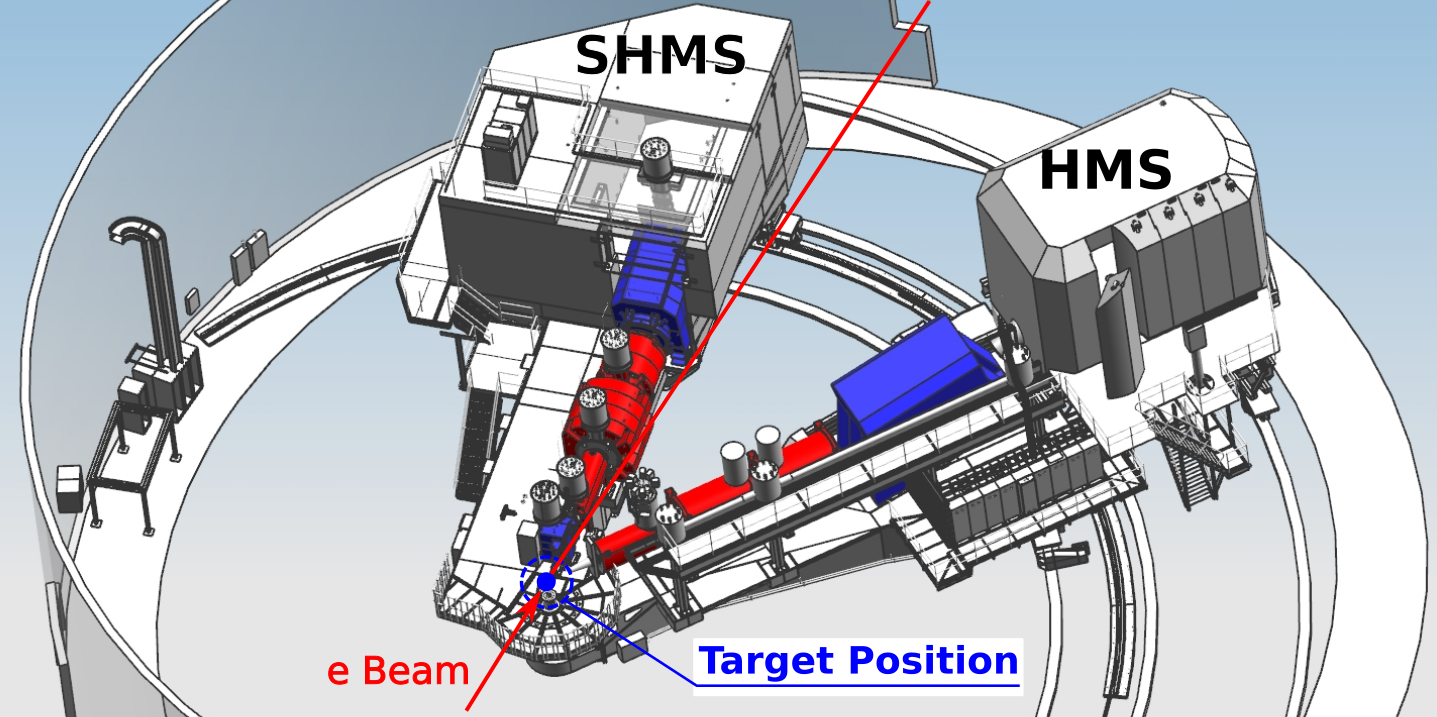}
 \caption[]{Hall C overview. The beam enters diagonally from lower left. In this example, the base equipment focusing spectrometers HMS and SHMS are configured for a coincidence (e,e$^\prime$ hadron) experiment.}  
  \label{fig:HallBaseSpec}
 \end{figure*}

\begin{figure*}
  \centering
\includegraphics[width=0.85\textwidth]{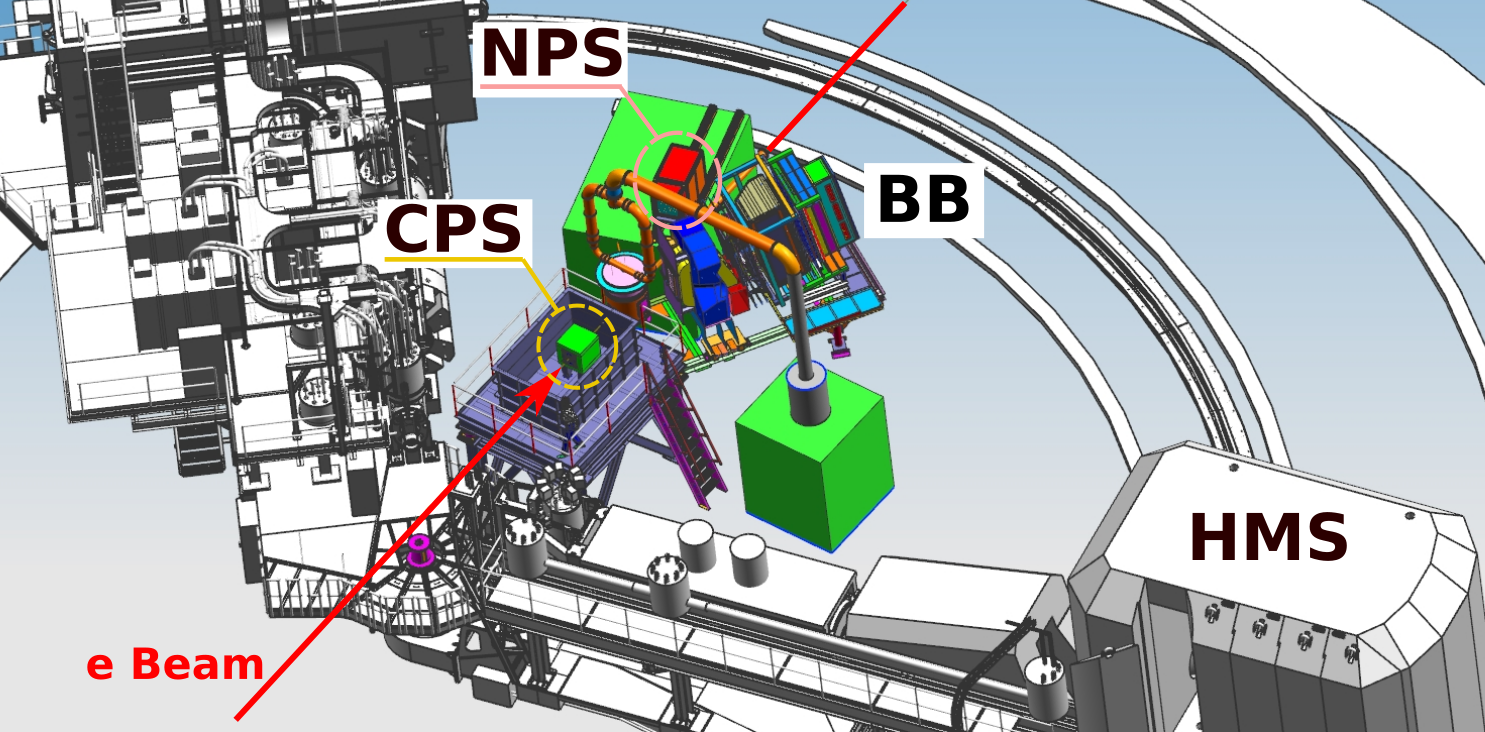}
\caption[]{ Here the base equipment focusing spectrometers are temporarily parked at large angles, making room for a proton Compton scattering experiment. Downstream of the standard pivot in this example are located the Compact Photon Source (CPS), a polarized proton target, and the large acceptance spectrometers Bigbite (BB) and the Neutral Particle Spectrometer (NPS). }
 \label{fig:HallwithCPS}
\end{figure*}

In the next decade, Hall C will be the only high luminosity, flexible, large-scale installation electron scattering facility in the world. Its base magnetic focusing spectrometers (HMS and SHMS), along with well-shielded detector huts and high power cryo-targets, allow routine operation at luminosities of $5 \times10^{38}$ $\textrm{cm}^{-2}\textrm{s}^{-1}$. Other equipment, such as the Neutral Particle Spectrometer (NPS), Compact Photon Source (CPS), multiple polarized targets, stand-alone lead-glass calorimeters, open detector magnetic spectrometers, and precision beam instrumentation are available for use in various combinations. As can be seen from examples in Fig. \ref{fig:HallBaseSpec} and \ref{fig:HallwithCPS}, Hall C provides floor space for a variety of experimental configurations utilizing these or other equipment. 
Moreover, as detailed in Section \ref{sect:precLT}, Hall C will be the only practical facility worldwide for precision Longitudinal-Transverse
(L-T)-separations at high $Q^2$ and $x$. This will enable the unique science program outlined in Table I below including, e.g., what are the spatial, momentum, and spin distributions of quarks and gluons inside the nucleon and nuclei?

In this paper, we will highlight the future program that is envisioned for Hall~C at Jefferson Lab, i.e. \emph{beyond} the already approved program of more than 5 years. We include an overview of additional opportunities for Hall C with the 12 GeV JLab featuring novel equipment and look forward into the next decade with possible accelerator upgrades. This paper is organized as follows.  After this Introduction in the present Section I, the following Section II summarizes the planned and possible physics program in Hall C. In Section III we overview capabilities of the existing Hall C facility including ongoing and envisioned upgrades to hall equipment. In Section IV we include a sampling of potential large experiments that could be carried out. In Section V, we discuss in a more detail two potential upgrades to the accelerator which could further expand the physics program.  A summary is provided in Section VI.

\section{Hall C Future Physics Program}
  
Hall C's outlook towards the future is delineated in Table~\ref{tab:hallc-future-science}. It depicts an active, leading-edge science program at 12 GeV, including the future possibilities of positron beams and higher beam energy. All these topics would take advantage of the hall's unique, high precision capabilities and versatility at the luminosity frontier.

\begin{table*}[hbt!]
\begin{centering}
\caption{Hall C at 12 GeV and future science} 

\footnotesize
\begin{tabular}{|p{2.3in}|p{1.9in}|p{2.3in}|}
\hline
{\bf Question} & {\bf Measurements} & {\bf Requirements} \\
\hline
What is the spatial distribution of quarks inside the nucleon and nuclei?
& Elastic and hard exclusive process cross-sections, parity violating DIS
&
High luminosity, excellent resolution, polarized beam and targets, CPS, positrons \\
\hline
What is the quark structure of nucleons and nuclei in momentum space?
& Inclusive and semi-inclusive cross-sections &
L-T separations, good resolution \\
\hline
How does the spin of the nucleon arise from the spin of quarks and their orbital angular momenta?
& $g_2$, $b_1$ & Transverse and tensor polarized targets \\
\hline
What is the nature of confinement/hadronization? 
& $(e,e'p)$ cross-sections, $A_{LT}$, SIDIS fragmentation functions, Hyperon decays
& Polarized beams, Focal Plane Polarimeter, low energy hadron detection, triple coincidence \\
\hline
What is the origin of hadronic mass? 
& Pion structure function via Sullivan process, & Low-energy forward hadron detection \\
  & $\pi^+$ and $K^+$ form factors & L-T separation, high momentum hadron detection at very forward angle \\
\hline
Where does the hard/soft QCD factorization regime begin? 
& Factorization tests in both $t$- and $u$-channel hard exclusive processes &
L-T separation, high momentum hadron detection at very forward and backward angles \\
\hline
What is the nature of the strong/nuclear force?
& Hypernuclear cross-sections, Short range correlations, Hyperon decays &
HMS, SHMS, NPS, polarized beam, backwards detection, triple coincidence \\
\hline

\end{tabular}

\label{tab:hallc-future-science}
\end{centering}
\end{table*}

Jefferson Lab and Hall C are the intensity frontier in the study of the quark-gluon structure
and the emergence of hadrons and the nuclear force from the color charge of QCD.
Studies of small cross sections and the ability to perform L-T separations,  both of which hinge on high precision,  will play a central role in the future program. The ability to install specialized detectors around the target also allows the detailed study of particular multi-particle final states in a high luminosity environment, especially when combined with new beamline facilities such as 
the CPS.

Since its commissioning in 1994, an enormous range of measurements have been performed in Hall C, 
from the weak charge of the proton~\cite{Nature:2018, Carlini:2019} in parity violating electron scattering using a low resolution, large acceptance spectrometer with  integrating-mode readout, to $\Lambda$-hypernuclear spectroscopy~\cite{Gogami:2021} based on high resolution spectrometers and excellent event-by-event PID. 
Precision form factors of the nucleon~\cite{Arrington:2003}, pion~\cite{horn06}, and kaon~\cite{Carmignotto:18} have been determined, 
predominantly via L-T separated cross sections. Nucleon and nuclear parton distributions, including transverse momentum dependence, have been measured, and the connection to resonance excitation of the nucleon explored~\cite{Stuart:1998}. This is by no means an exhaustive list, but serves to demonstrate the capabilities and flexibility of Hall~C. 
This broad range will carry into the future with an additional focus on the exclusive reactions which uncover the intricacies of the generalized parton distributions (GPDs)~\cite{Ji:1997, Diehl:2003}.


L-T separations are essential to the Hall C scientific program, and the technique is utilized in a wide variety of exclusive, semi-inclusive, and inclusive measurements. The separation of the cross-section into its components simplifies the theoretical interpretation of the experimental measurement and tests our understanding of how QCD processes factorize into hard and soft physics distributions. In the near future, it will be applied to the relatively unexplored $u$-channel kinematical domain, probing nucleon structure across the entire impact parameter space from $|t_{\textrm{min}}|$ to $|t_{\textrm{max}}|$.

 


In the remainder of the present section, we will very briefly discuss the physics topics in Table~\ref{tab:hallc-future-science}. When applicable, we will refer to a sampling of potential experiments in section IV.  

Hard, exclusive reactions, at both low and high momentum transfer $-t$, will play an important role in probing the three-dimensional structure of the nucleon via GPDs and Transition Distribution Amplitudes (TDAs). Examples of hard, exclusive  reactions include elastic scattering, Deeply Virtual Compton Scattering (DVCS), Time-like Compton Scattering (TCS), and Deeply Exclusive Meson Production (DEMP). The large space- or time-like mass of one or more virtual photons in these reactions leads to very small cross sections which therefore require high luminosity for precision measurements.  

Semi-inclusive reactions will be used to explore details of the quark momentum distributions using flavor tagging to uncover details of the flavor and large transverse momentum dependence. 
Both exclusive and inclusive measurements are discussed in subsection IV.A and will help determine how the spin of the nucleon arises from its constituents as described in IV.D. These measurements rely on the availability of high performance, polarized proton, deuteron, and $^3\text{He}$ targets. We note that vector and tensor polarized deuterium targets will likely remain a unique capability of Jefferson Lab for many years. More general studies of the nature of confinement and hadronization are described in IV.B. 

Color transparency studies in nuclei with $(e,e^{\prime} p)$, as well as the asymmetry $A_{LT}$, provide insight into the very early stages of how the struck point-like configuration (PLC) transforms into a hadron.  Focused studies of fragmentation looking at particle correlations and transverse momentum dependence are also foreseen. As described in sections IV.F.2 and IV.H, there are other observables which probe this physics in entirely new ways at Jefferson Lab energies. It has been realized for some time now that understanding the quark structure of light mesons 
addresses the question of the origin of hadronic mass, one of the most important questions in QCD physics. Both pions and kaons can be explored in Hall C, as described in IV.F.1. 
 
The nuclear force between strange and non-strange baryons can be explored via the spectroscopy of $\Lambda$ hypernuclei. 
Precise spectroscopy of hypernuclei is one of key studies to provide information on the 
three-body baryon repulsive force which could make neutron stars’ equation of state stiff enough to support heavy
neutron stars. Experimental configurations with the necessary intensity, resolution, and particle identification capabilities are outlined in IV.C .

The strange quark mass is closest to the QCD cut-off scale, a fact which has long been recognized as the key to unlocking the mystery of QCD confinement. Section IV.H describes a unique program to study confinement in QCD with precision measurements of self-analyzing hyperon decays. This program relies on Hall C's unique capability to study multi-particle final states in a high luminosity environment.
 



Generally speaking, the scope of all the programs above would be increased with higher beam energy, while the availability of positrons would open new and complementary areas of study.
With an energy upgrade increasing the center of mass energy up to 7 GeV, Hall C would be able to access additional physics topics such as Double DVCS, Deeply Virtual Meson Production in the scaling region, SIDIS kaon ratios such as $K^+/K^-$ on both proton and deuteron targets, EMC and anti-shadowing, PV elastic scattering to constrain strange quark form factors, and also considerably increase the kinematic reach for precision cross sections, e.g., for pion and kaon form factors, SIDIS basic cross sections, tagged DIS, $x>1$ and EMC effect, threshold charm states production, and (e,e$^{\prime}$) backward particles.
A positron beam would provide opportunities to measure the interference amplitudes from two-photon exchange as well as search for BSM physics via, e.g., dark photon searches. A positron-electron source may be accelerator based or, as described below in Section \ref{CPS_e+}, created inside Hall C using the CPS photon source with a downstream target for pair production.

\section{Overview of the Hall C facility}

In this section, we overview the existing Hall C capabilities as well as those that will become available in the near future. 
A few delineated sections refer to either a significant Hall C detector upgrade or a significant accelerator upgrade. 
It is critical to note (a) the precision measurements made possible by this equipment, (b) the flexibility of configurations it affords, and (c) the generally high rate and/or high luminosity of which it is capable. These three key points provide Hall C with a unique and critical role in the coming scientific era. 
 
\subsection{Base Focusing Spectrometers}


%
The base detectors for Hall C are the High Momentum Spectrometer (HMS)
and Super High Momentum Spectrometer (SHMS).  These moderate acceptance (several msr) devices can reach momenta of 6+ $\textrm{GeV}/c$ and 11 $\textrm{GeV}/c$, respectively,  with momentum bites exceeding 10\%. 
Momentum reconstruction resolution and accuracy at the 0.1\% level are routinely achieved. Generally excellent particle identification is achieved by a combination of Cherenkov detectors and lead-glass shower counters. 
These spectrometers can be placed at scattering angles as small as 5.5$^{\circ}$ (SHMS)
and 10.5$^{\circ}$ (HMS) with sub-mrad pointing accuracy. They may be used in single-arm or in
coincidence mode, or with other detectors, and have well shielded
detector stacks.  This provides a unique ability to measure small
cross sections which demand high luminosity while allowing the
careful study of systematic uncertainties.
Additional information about the Hall C base spectrometers can be found in the Hall C Standard Equipment manual. \cite{HallC_SEM}
 
When the magnetic spectrometers are rotated to large angles, the 13$'$ beam height provides the flexibility to install a
variety of additional detectors.  Historically, this flexibility
has allowed for large acceptance devices for parity violating
electron scattering, extremely high resolution focusing spectrometers optimized for hypernuclear physics, 
and other equipment. In the future, it can accommodate devices such as
the Super Bigbite and Bigbite (SBS and BB) discussed in Section III.G, the large hadronic calorimeter HCAL or electromagnetic calorimeter ECAL discussed in Section III.H, or other new detectors.

\subsection{Targets}


A variety of targets 
are routinely available in Hall C.
The standard target assembly allows for liquid
hydrogen/deuterium or dense ${}^3\textrm{He}$/${}^4\textrm{He}$ all of
which may be operated at high luminosities.  Other targets have
included dynamically polarized hydrogen and deuterium targets
(in the form of $\textrm{NH}_3$ and $\textrm{ND}_3$) as well as polarized
${}^3\textrm{He}$.  
In the near future, a combination of longitudinally, transversely, and tensor polarized ammonia targets will also be available.

\subsubsection{High Power Targets}

\begin{figure}[hbt!]
  \centering
\includegraphics[width=0.48\textwidth]{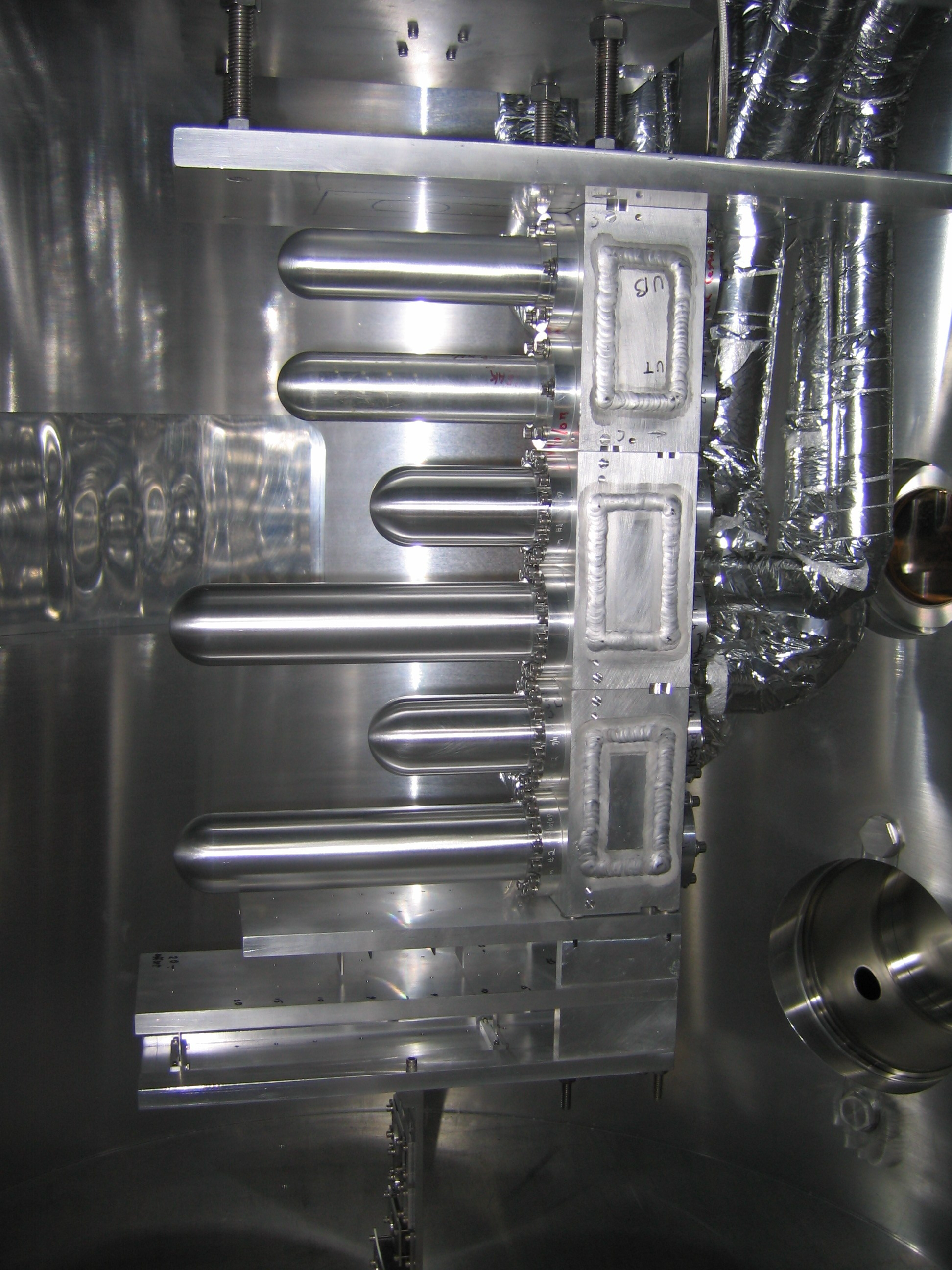}
 \caption[]{ 
 The cryotarget configuration. Each independent loop has 2 cell lengths available. Cell lengths of 4cm, 10cm, and 15cm appear in this photo. The solid target ladder is suspended from the bottom. 
 }
  \label{fig:CryoTgt}
 \end{figure}

High power cryo-targets are critical to many Hall C experiments. 
The standard stack of cryo-targets system is  comprised of three independent, closed circulation loops. Typically, one loop contains liquid hydrogen  ($\textrm{LH}_2$), one contains liquid deuterium ($\textrm{LD}_2$), and one is a spare. Each cryogenic loop can have one or two cells stacked on top of each other as seen in Fig.  \ref{fig:CryoTgt}. There is a single-axis vertical motion system with a range of 60 cm for positioning a target on the beam axis. 

The cryotarget cells which have been used most in recent years are 10 cm in length and operate with beam currents up to $100~\mu\textrm{A}$ (for a maximum luminosity for $\textrm{LH}_2$ of $3 \times 10^{38}~\textrm{cm}^{-2} \textrm{s}^{-1}$). 
Because of the high internal flow rate and fast beam raster, the density reduction in the beam path is at most several percent. Monitoring of the temperature and pressure of the fluid permits the target density to be determined with a precision of 1\% or better.  

With appropriate planning and funding, even higher power (or otherwise specialized)
targets can be employed. For example, the $Q_{\textrm{weak}}$ experiment completed running
on its $\textrm{LH}_2$ target in Hall C in 2012. At the time, the Q$_{\textrm{weak}}$
target was the highest power $\textrm{LH}_2$ target in the world at 2500 W.
This was the first target at Jefferson Lab designed with computational fluid dynamics
(CFD). The excellent performance of the $Q_{\textrm{weak}}$ target validated CFD as a
critical tool in designing high power, low noise targets.\cite{QweakBeamline}
%
%

Building on that success, in 2016 a CFD facility (CFDFAC) was established at Jefferson Lab with funding from a DOE Early Career Award. The CFDFAC comprises a high performance computing (HPC) homogenous farm with 256 physical CPUs. The software used is ANSYS-CFD, which includes both Fluent and CFX. The CFDFAC has been tasked with designing high power, low noise targets for the 12 GeV program at Jefferson. The cells designed  
show an improvement of an order of magnitude in performance compared with similar length cells used during the 6 GeV program. CFDFAC is now a general-purpose facility involved in designing and assessing the thermodynamic performance of targets at Jefferson Lab.

High density ${}^{3,4}\textrm{He}$ targets require thicker windows than hydrogen cells because they must operate at high pressure (200 psi at 5K). A new 10 cm long cell design will be deployed in 2022 which has 100\% transverse flow to help minimize density reduction in the beam path. 

Many solid targets are available as physics targets, for background studies, and for spectrometer optics studies.
The solid target ladder is conductively cooled via connection to the underside of the stack of cryotarget blocks.   The maximum beam current to which a solid target can be subjected depends on the specific material and how well it is conductively coupled to the cryotarget block.
For example, a 5\% carbon target has been operated at $60~\mu\textrm{A}$ to achieve an electron-nucleon luminosity of $4.8 \times 10^{38}~\textrm{cm}^{-2}\textrm{s}^{-1}$.
Also, a 1.5\% aluminum alloy target is frequently used for target window background subtraction. This is called a ``window dummy'' target. It consists of two separated foils, carefully matched to the 10 cm long cryotarget cell in terms of the aluminum alloy used,  the total radiation length, and the 10 cm window separation. 

Cryotarget cells and a window dummy target with 4 cm length are also available and could be useful in systematic studies or structure function measurements where precise knowledge of the acceptance is necessary.  The trade-offs between acceptance and window backgrounds and radiative corrections can be explored. However, the generally low counting rate of high $Q^2$ measurements in the 12 GeV era has meant that most spokespersons have preferred the 10 cm cells. 

An extremely intense beam of photons can be produced with a
target-like Bremsstrahlung radiator placed upstream of the target of interest.
Such a configuration has been used for very low 
cross section measurements such as $d(\gamma,p)n$ and for J/Psi photoproduction on the proton. 
The resulting mixed photon/electron beam is not ideal for a solid polarized target, however, since the electron component in the beam heats and reduces the polarization of these targets thus limiting the deliverable photon flux. 
Pure photon beams can be produced with the new Compact
Photon Source (CPS) discussed in Section III.D . 

In summary, the package of liquid cryogenic and solid targets discussed in this section is well-suited for precision measurements at high luminosity. Some important target parameters are summarized in Table \ref{tab:cryotarget}. 

\begin{table}[ht]
\caption{Parameters of several example high power targets.}
\label{tab:cryotarget}
\centering
\begin{tabular}{ccc}
Target              &  Thickness                                   &  Max. Current              \\
                    &                                                      &   (Power)          \\  \hline
$\mathrm{LH_2}$, $\mathrm{LD_2}$        &  10 cm                                      &   $100~\mu\textrm{A}$  \\   
                    &    1.5\% RL                                 &    (400 W)                  \\  \hline
$^{3}\textrm{He}$   &  10 cm                                         &   TBD                          \\
                    &($0.105~\textrm{g}/\textrm{cm}^3$)              &                                \\  \hline 
$^{4}\textrm{He}$   &  10 cm                                         &   TBD                            \\
                    & ($0.140~\textrm{g}/\textrm{cm}^3$)             &                                \\   \hline
carbon              &    5\% RL                                     &    $60~\mu\textrm{A}$     \\
                     &                                               &   (380 W)                 \\  \hline
aluminum            &    1.5\% RL                                    &   $40~\mu\textrm{A}$      \\
                     &                                               &      (~32 W)               \\
\hline

\end{tabular}
\end{table}

 
\subsubsection{Polarized Targets}\label{sect:poltarg}


Hall C utilizes dynamically polarized $\text{NH}_3$ and $\text{ND}_3$ solid targets with a 5 T superconducting magnet and 1 K helium evaporation refrigerator.  This system provides proton vector polarization greater than 90\%, deuteron vector polarization greater than 40\%, and operates with beam currents up to 140 nA providing a total luminosity approaching $10^{36}~e$-nucleon $\textrm{cm}^{-2}\textrm{s}^{-1}$. 

The split-coil magnet can be rotated to produce a horizontal magnetic field that is either parallel to the incoming beam (longitudinal polarization) or perpendicular to the beam (transverse polarization). 
For future experiments focused on transverse polarization, a new 5 T magnet with a wider split has been procured that increases the transverse coverage to $\pm$25 degrees while retaining $\pm$35 degrees longitudinal coverage.  The new magnet can also be rotated to produce a vertical field, thus giving a target polarization that is 90 degrees out-of-plane from the electron beam and Hall C spectrometers. The new magnet does not require liquid helium for operation and is compatible with the other components of the polarized target system.

An R\&D program is in progress to increase the deuteron tensor polarization from a current maximum value of about 15\% to more than 30\% by RF saturation of selected areas of the deuteron's NMR lineshape.  The RF is used to depopulate the m=0 substate of the deuteron spin, thus increasing the tensor polarization, $P_{zz} = 1 - 3N_{0}$.  The resulting tensor polarization can be determined using a sophisticated analysis of the NMR lineshape, which in turn can be benchmarked against known values of the tensor analyzing power $A_{zz}$ at low energies. 

\begin{figure}[hbt]
  \centering
  \includegraphics[width=0.6\textwidth]{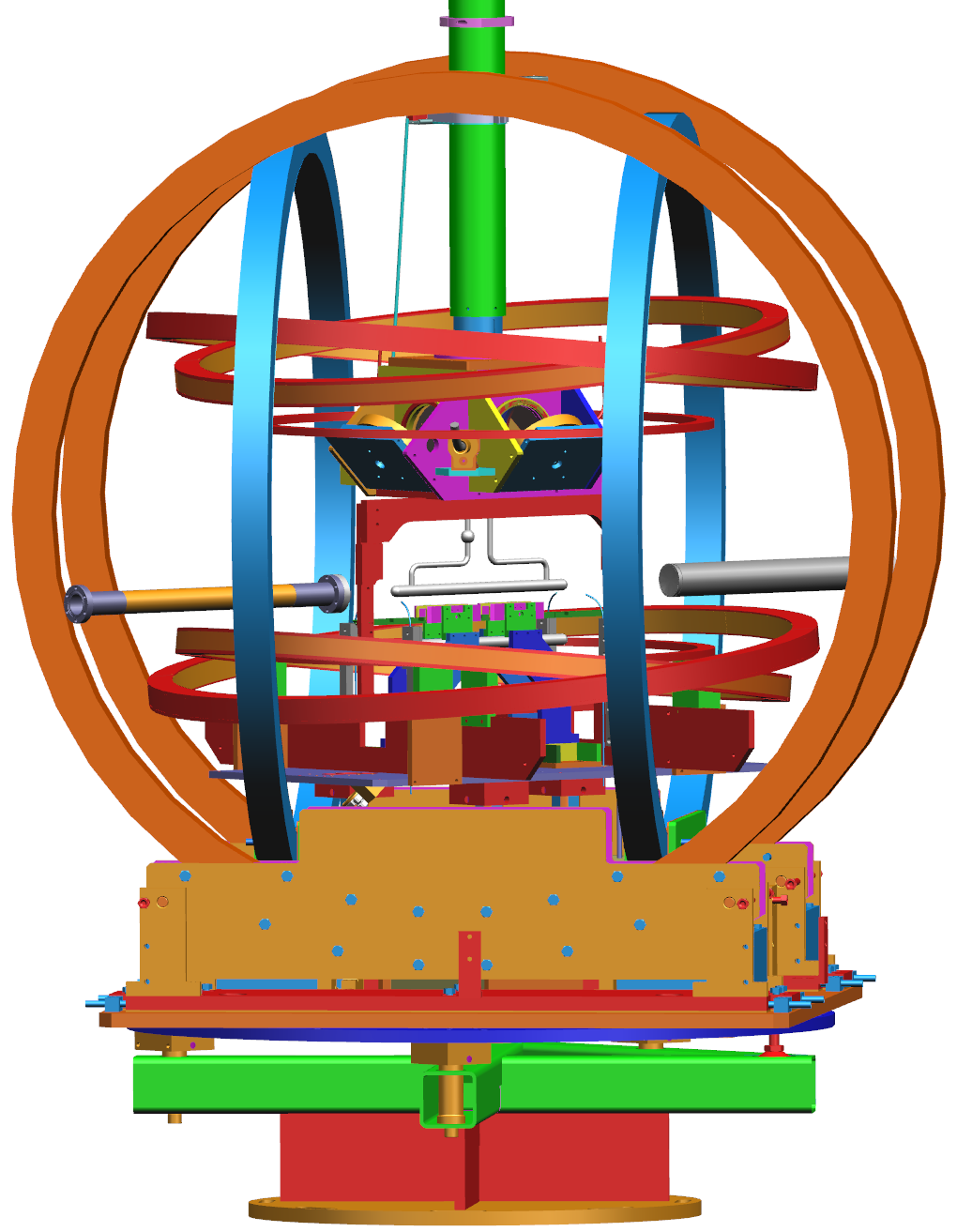}
  \caption[]{CAD rendering of a polarized $^3$He target configuration used in
     Hall C.  The $^3$He target cell is the horizontal grey cylinder in the
     center of the frame.  The large rings oriented vertically and horizontally
     make up three independent pairs of Helmholtz coils which provide polarizing magnetic fields in arbitrary direction.  Laser power for optical
     pumping of the target is directed from the top of the frame to the pumping
     chamber above the target cell. 
 }
  \label{fig:polhe3}
 \end{figure}

The JLab polarized $^3$He target in Fig.~\ref{fig:polhe3}
is based on optical pumping using high
performance, high power, narrow-width diode lasers and Rb-K hybrid
spin-exchange. For the recent Hall C $A_1^n$ and $d_2^n$ experiments
\cite{E12-06-110, E12-06-121}, this
target achieved world-record performance in terms of
figure-of-merit. With a 30\,uA electron beam on a 40\,cm long, 10\,amg
target, the luminosity is $2.2 \times 10^{36}~e$-nucleon $\textrm{cm}^{-2}\textrm{s}^{-1}$. 
The use of convective gas flow, first introduced in Ref.~\cite{Dolph:2011rc}, allows the target in-beam polarization to reach over $55\%$.
The target can be polarized in longitudinal, transverse, vertical, or any
direction with 3 pairs of Helmholtz coils. The target
spin can be flipped every few minutes if needed, allowing one to
perform precision measurements of target single spin asymmetries.
The ongoing upgrade for the upcoming GEn-II target aims to increase the
luminosity to $6.6 \times 10^{36}$ e-nucleon $\textrm{cm}^{-2}\textrm{s}^{-1}$ with an in-beam
polarization of up to $60\%$. A separate R\&D effort \cite{Anderson:2020hlt} aiming to
reach a luminosity into the mid-$10^{37}$ $e$-nucleon $\textrm{cm}^{-2}\textrm{s}^{-1}$ range is also under consideration.

The physics enabled by these two polarized targets falls into several broad
categories, which we discuss in section \ref{poltarphysics} below. 

\subsection{Electron Beam Polarimetry}


\subsubsection{Beam Polarimetry in the 12 GeV Era}

Experimental Hall C uses two techniques for measuring electron beam polarization: M\o ller and
Compton polarimetry.\footnote{See Ref.~\cite{Aulenbacher:2018weg} for a detailed description of these
techniques.}
M\o ller polarimetry makes use of the analyzing power in the scattering of polarized beam electrons
from atomic electrons polarized by an applied magnetic field in a ferromagnetic foil target. While
M\o ller polarimetry typically benefits from relatively short measurement times, the measurements are destructive to the electron beam and can only be carried out at low currents to avoid target depolarization due to heating. Compton polarimeters use the scattering of circularly
polarized laser light from polarized beam electrons.  Compton measurements are essentially non-destructive with respect to the beam properties, so can
be performed concurrently with beam delivery to the running experiment.  Measurement times can be
longer, making precise systematic studies difficult to schedule.  In addition, the analyzing power varies
rapidly with beam energy, complicating the design of a device capable of making measurements over a broad range of beam energies.
Having both polarimetry techniques available as cross-checks on one another has been extremely valuable in identifying and reducing systematic errors. 

\begin{figure*}[htb]
  \begin{center}
    \includegraphics[width=\linewidth]{ 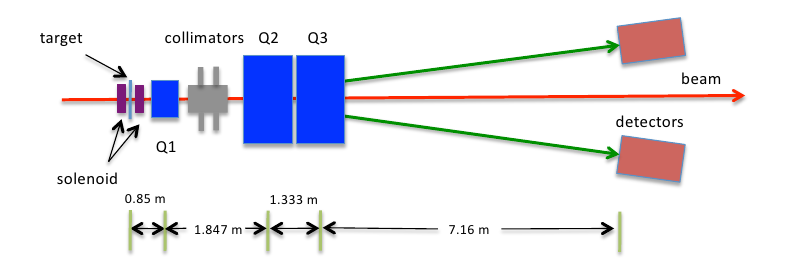}
    \caption{Layout of the Hall C M\o ller polarimeter.  A pure iron foil is polarized out-of-plane
      in a high field, split-coil solenoid magnet while a quadrupole-based magneto-optical system
      (Q1, Q2, Q3) focuses the scattered and recoiling electrons on a pair of lead glass calorimeters 
      about 11~m downstream.} \label{fig:moller}
  \end{center}
\end{figure*}

The Hall C M\o ller polarimeter, shown in Fig.~\ref{fig:moller}, was the first to make use of a target system consisting of a pure iron foil driven to
magnetic saturation via a high magnetic field (3-4 T) normal to the foil. 
Iron was chosen due to its well-understood magnetic properties. 
This technique reduced what had historically been the dominant source of systematic uncertainty (knowledge of the target polarization)
from typical values of $\approx$2-3\% to $\approx$0.25\%~\cite{bever:1997kvc}, allowing total systematic uncertainties on the order of 0.5\%~\cite{Hauger:1999iv}.  As originally constructed and
used during the 6 GeV program, the Hall C M\o ller made use of a 2-quadrupole magneto-optical system
to focus the scattered and recoil electrons onto a pair of lead glass calorimeters, where they were detected in
coincidence with very low background.  In order to accommodate operation at 11 GeV, an additional large quadrupole (Q3) was
added to the system.  Q3 is wired in series with Q2 so that the pair effectively function as one magnet
with twice the effective length.  Additional modifications were also required to the
background-reducing collimators between Q1 and Q2 due to the smaller lab scattering angle at higher
beam energy.  The Hall C M\o ller was successfully operated in this new configuration in 2019 and 2020
during the running of the Hall C polarized $^3$He target program.

\begin{figure*}[htb]
  \begin{center}
    \includegraphics[height=\linewidth,angle=-90]{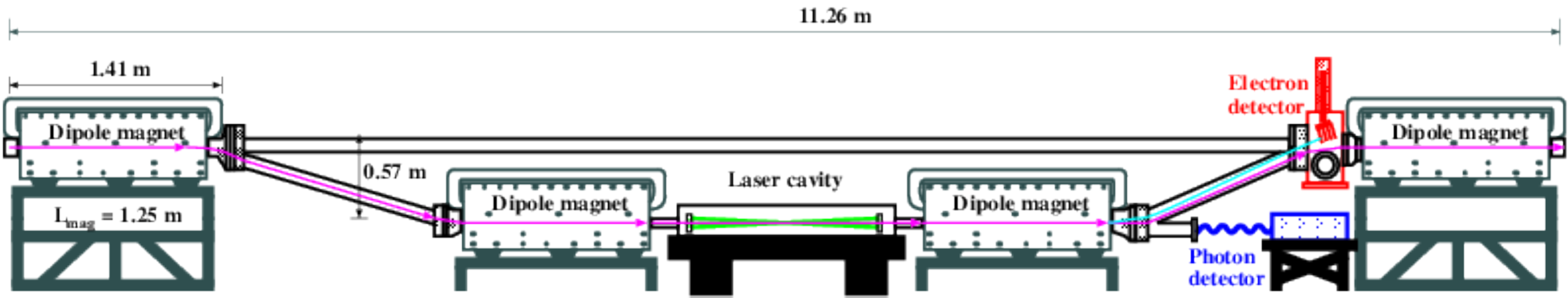}
    \caption{Layout of the Hall C Compton polarimeter as originally constructed in 2010.  A
      4-dipole chicane deflects the electron beam vertically to interact with a high-power laser beam
      between dipoles 2-3, and restores
      the electron beam to its nominal height via dipoles 3-4.  Backscattered photons are detected in a lead tungstate
      calorimeter, while the scattered electrons are detected in a position-sensitive diamond strip detector just upstream of the
      4th dipole (figure from~\cite{Narayan:2015aua}).  For operation at 11 GeV, the vertical deflection
      of the electron beam was reduced from 57 cm to 13 cm.}\label{fig:compton}
  \end{center}
\end{figure*}

The Hall C Compton polarimeter was built and installed in 2010, just prior to the start of a parity-violation experiment to measure the weak charge of the proton (Q$_{\textrm{weak}}$). Its design is based in large part on the successful Hall A Compton polarimeter, incorporating a laser system utilizing an external Fabry-P\'{e}rot cavity
positioned at the center of a four-dipole chicane as shown in Fig.~\ref{fig:compton}.  Backscattered photons
are detected in a lead tungstate calorimeter downstream of the laser, while scattered electrons are
momentum-analyzed in the third chicane dipole and detected in a 4-plane diamond strip detector.  

The Hall C Compton was commissioned in 2010-2011 and operated very successfully in 2011-2012, resulting
in a final systematic uncertainty of $\frac{\Delta P}{P}=0.59\%$~\cite{Narayan:2015aua} for that run period.
During the same period, a direct comparison of the Hall C Compton and M\o ller polarimeters was
performed by operating both devices consecutively at the same beam current.
The two polarimeters were found to agree within their respective
uncertainties. By varying the beam current, limits were placed on the possible dependence of beam polarization on current~\cite{Magee:2016xqx}.

For operation up to 11 GeV beam energies, the Compton polarimeter has since been modified by reducing the size of the vertical deflection of the electron beam in the 4-dipole
chicane from 57~cm to 13~cm (given by the field constraints imposed by the dipoles).  This large
reduction of the vertical deflection required significant modifications to the vacuum system as well as 
new, larger aperture corrector magnets. 

\subsubsection{Polarimetry Outlook with Positron Beams}
An exciting program of experiments using polarized positron beams has been proposed at Jefferson Lab~\cite{EPJApositrons}.
Polarimetry will be a key requirement for these measurements.  Fortunately, the Hall C polarimeters should
be able to measure positron polarization, although with some updates and modifications~\cite{Gaskell:2018hth}.
The Hall C Compton polarimeter is, in principle, capable of measuring positron beam polarization with no changes, other than
flipping the polarity of the dipole chicane magnets.  The expected relatively low positron beam current (on the order of 100 nA) and polarization
(>50\%), however, will have significant impact on the measurement time.  At 11 GeV, the Hall C Compton
would be able to make a $\frac{\Delta P}{P}=1\%$ measurement in about 9 hours, assuming no changes to the system. More
rapid measurements could be made by updating the laser system to use an RF-pulsed, mode-locked laser 
coupled
to the external Fabry-P\'{e}rot cavity.  Assuming the same average laser power, a pulsed system would result in
about a factor of 10-15 improvement in luminosity, reducing the time needed to achieve 1\% statistical errors to O(1) hour.

The M\o ller polarimeter can also be used to measure positron beam polarization, but significant
modifications would be required.  
The lower positron beam current can be overcome by simply using thicker targets (10-20~$\mu$m foils as opposed
to 1-4~$\mu$m).  However, steering the scattered and recoiling particles
to the existing M\o ller detectors becomes problematic.  The M\o ller optical system is designed for focusing particles of the same
charge.  For positron beams, we must detect the scattered positron and recoiling electron in coincidence.  This will
require replacing the quadrupole-based optical system with one based on a (large gap) dipole.  Simple simulations show that
a dipole with a field integral of 1 T-m would be able to focus the positron-electron pair to the detector plane at 11 GeV.
It is possible that this could even be achieved by re-wiring one of the large M\o ller quads (Q3) to provide that dipole field.\footnote{Such an approach is being pursued in Hall B.} This would leave both the smaller quad (Q1)
and the remaining large quad (Q2) for beamline optics, however more detailed simulations of the magnets are required.

\subsubsection{Polarimetry Outlook at 20 GeV}

Given the additional space along beamline required for optimal operation of the Compton and M\o ller polarimeters at higher energy,
it is possible that a higher energy Hall C program would require the removal of either the Compton or M\o ller to
free up the space needed to have at least one high precision polarimeter.
A detailed discussion of the trade-offs can be found in the Appendix.


\subsection{Compact Photon Source (CPS)}

%
%
The CPS is designed to provide a high intensity and narrow Bremsstrahlung beam for experiments involving exclusive, photon-induced reactions. As with a simple Bremsstrahlung raditor, if the primary electron beam is longitudinally polarized, then photons near the energy endpoint are highly circularly polarized. 
But unlike a simple Bremsstrahlung radiator, the CPS photon beam will be  essentially free of electrons and their ionization heating, hence it is extremely well suited for use with polarized targets. The CPS will enable 
measurements of photon-nucleon interactions at both high $s$ and high $-t$ such as WACS (Wide Angle Compton Scattering) which employs a transversely polarized target, as well as exclusive reactions with a focus on small cross sections/exclusive photoproduction. Two examples of the latter are E12-17-008 and the Timelike Compton Scattering proposal (C12-18-005).  
 
\begin{figure}[hbt!]
  \centering
\includegraphics[width=0.6\textwidth]{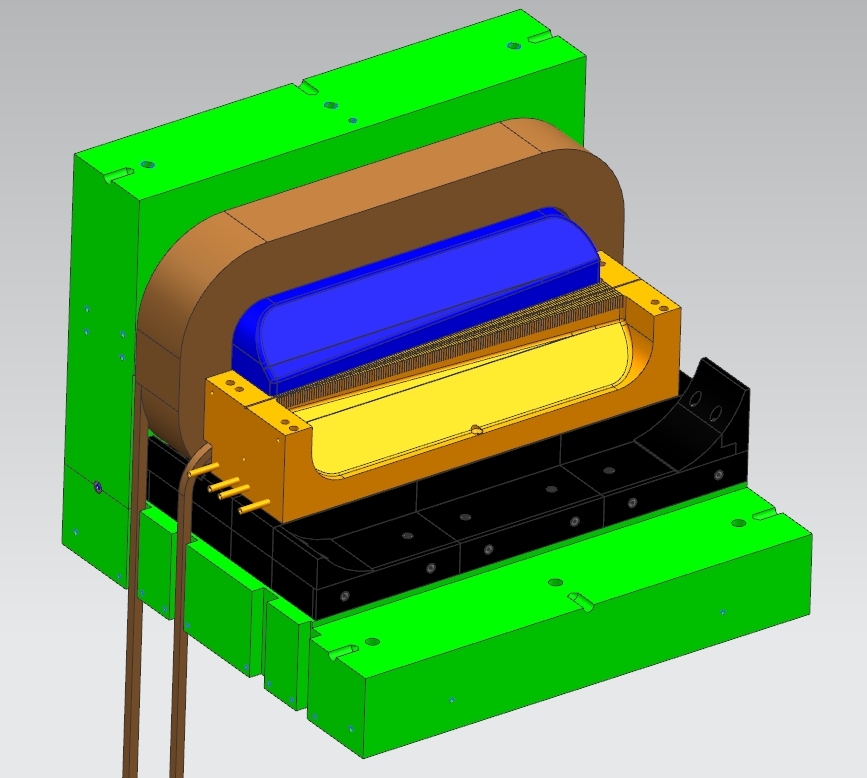}
 \caption[]{A CAD rendition of the Compact Photon Source (CPS) without its external shielding and part of yoke removed for clarity. From inside to outside: the slotted Cu power absorber is shown in yellow; the magnet pole is blue; the WCu inner shielding is black, the magnet coil is brown, and the magnet yoke is green.   
 }
  \label{fig:cps-magnet}
 \end{figure}   

The CPS consists of a 10\% radiator located upstream of a dipole magnet to deflect the primary electron beam into a central copper absorber surrounded by tungsten powder and borated plastic for shielding. (See Fig.  \ref{fig:cps-magnet}.) Sharing elements with both a traditional photon tagger and conventional relatively high-power (30kW) beam dump, its design is optimized to simultaneously produce a narrow ($\sim$1 mm) photon beam and provide effective shielding with an overall compact geometry~\cite{Day:2019qdz, BW2017}. Detailed Monte Carlo studies guided the technical design, taking into account power deposition in the CPS and the shielding of prompt and residual activation.

It is anticipated that the CPS will be able to operate with an 11 GeV electron beam up to a maximum current of $2.7~\mu \textrm{A}$ (30kW), leading to a secondary beam intensity of $10^{12}$ equivalent quanta per second. This represents a factor of $\sim$30 increase over traditional techniques of photon beam production. The intensity of the CPS photon beam is sufficiently high that its use as a secondary beam source is also being explored.  One such exciting opportunity in Hall C is the possibility of using it to drive a broad-band positron source. As discussed in Section~\ref{CPS_e+}, this concept could be utilized in new, high-precision studies of the two-photon exchange amplitude.  In addition, there are plans to use a CPS-like design in experimental Hall D to produce a $K_{\textrm{long}}$ beam.

\subsection{Hall C Beamline and Diagnostics} 

The Hall C beamline is divided into two parts by a shield wall. Upstream of the wall is the accelerator enclosure where the arc magnets reside, whereas downstream of the wall are the final focusing magnets, polarimeters, and fast raster.  
Safety systems permit personnel to work in Hall C while beam is being delivered to the other halls.  

\subsubsection{Optics and Fast Raster}
 
The Hall C beamline after the shield wall was redesigned for the $Q_{\textrm{weak}}$ experiment \cite{QweakBeamline} with consideration for the subsequent beam energy increase to 11 GeV. 
The beamline contains Compton and Moller polarimeters, with beam waists at both the Compton interaction region and the target. The latter waist can be moved downstream of the standard pivot to accommodate experiments with special downstream target configurations.  

The fast target raster is located after the last quadrupole in order to ensure that the raster pattern on the target (nominally 2 mm $\times$ 2 mm) is independent of beam optics.
The iron-free raster magnet coils are driven by a triangular wave to improve the raster uniformity. The fundamental frequency is near 23 kHz, with X and Y driven at slightly different frequencies to minimize dwelling patterns on the targets.  

\subsubsection{Beam Diagnostics}

The Hall C beamline beam position monitors (BPMs) are specified to have an accuracy of 100 $\mu$m at 1 $\mu$A of beam current for 1 second integration time. 
Internally, there are 4 antennae, and the beam position is proportional to the difference over the sum of the signals induced in a pair of opposite antennae. 
Because they use JLab's sophisticated switched-mode electronics,  their offsets are highly stable. The BPMs used in Hall C can operate down to beam currents as low as 50 nA before losing lock, which is valuable for polarized ammonia target operations. 

Low noise measurements of the beam current with long term stability at the 1\% level are made using five cylindrical RF  cavities. These so-called beam current monitors (BCMs) are tuned to resonate in $\mathrm{TM_{010}}$ mode at 1.497 GHz. For absolute calibration of the BCMs, an Unser monitor is used. The latter is a toroidal parametric current transformer which is rather noisy at $\mathcal{O}(1)$ $\mathrm{\mu A/\sqrt{Hz}}$ but has excellent long term gain stability. The Unser gain can be readily checked at the 0.1\% level by injecting a precision current through a wire passing  through the toroid.  

Just upstream of the standard target pivot, there is a 240 cm long diagnostic girder including BPMs and wire scanners (to measure the beam profile). Orbit locks ensure that the beam position on target is stable to 100 $\mu$m.
Immediately upstream of the target scattering chamber is a final BCM which is used by the Machine Protection System (MPS) to ensure that minimal beam current has been lost between the injector and the experimental halls. 
 
 \subsubsection{Beam Energy Measurement System}
The beam energy is measured using a string of upstream dipoles. The field integrals of the dipoles are known to better than 0.1\%.  During a measurement, the quadrupoles and corrector dipoles are turned off, so that electron beam energy can be determined from the field integral and a precise measurement of the net bend angle into the hall.

 To make the angle measurement there are two granite benches, one just before the first dipole, and one just after the last dipole and shield wall.  Each granite bench has two precision wire harps separated by 250 cm whose absolute and relative locations are known by survey.  These two pairs of harps are used to determine the deviations from the nominal 34.3$^{\circ}$ bend into Hall C.


\subsection{Neutral Particle Spectrometer (NPS)  }


%
%
The NPS is an example of key new equipment that will become available in Hall C in the next few years. The NPS will give access to precision measurements of small cross sections for reactions with photons or $\text{e}^{+/-}$ in the final state. It consists of a high resolution,  $\text{PbWO}_4$, electromagnetic calorimeter, and its science program currently features six fully approved experiments which demonstrate the importance of precision cross section capability. The E12-13-010 and E12-06-114 experiments will measure the Exclusive Deeply Virtual Compton Scattering and $\pi^0$ cross sections to the highest $Q^2$ accessible at Jefferson Lab. Both experiments will provide important information for understanding Generalized Parton Distributions (GPDs). The E12-13-007 experiment will study the semi-inclusive $\pi^0$ electroproduction process and seeks to validate the factorization framework that is needed by the entire 12 GeV semi-inclusive deep-inelastic scattering program and beyond. Measurements of Wide-Angle (WACS) and Timelike Compton Scattering (TCS) reactions will be performed by the E12-14-003, E12-17-008 and C12-18-005 experiments. These measurements will allow for a rigorous test of the universality of GPDs using high-energy photon beams. The NPS will also be used in the E12-14-005 experiment to study exclusive production of the $\pi^0$ at large momentum transfers in the process $\gamma p \rightarrow \pi^0 p$. 

\begin{figure*}[hbt!]
  \centering
\includegraphics[width=1.0\textwidth]{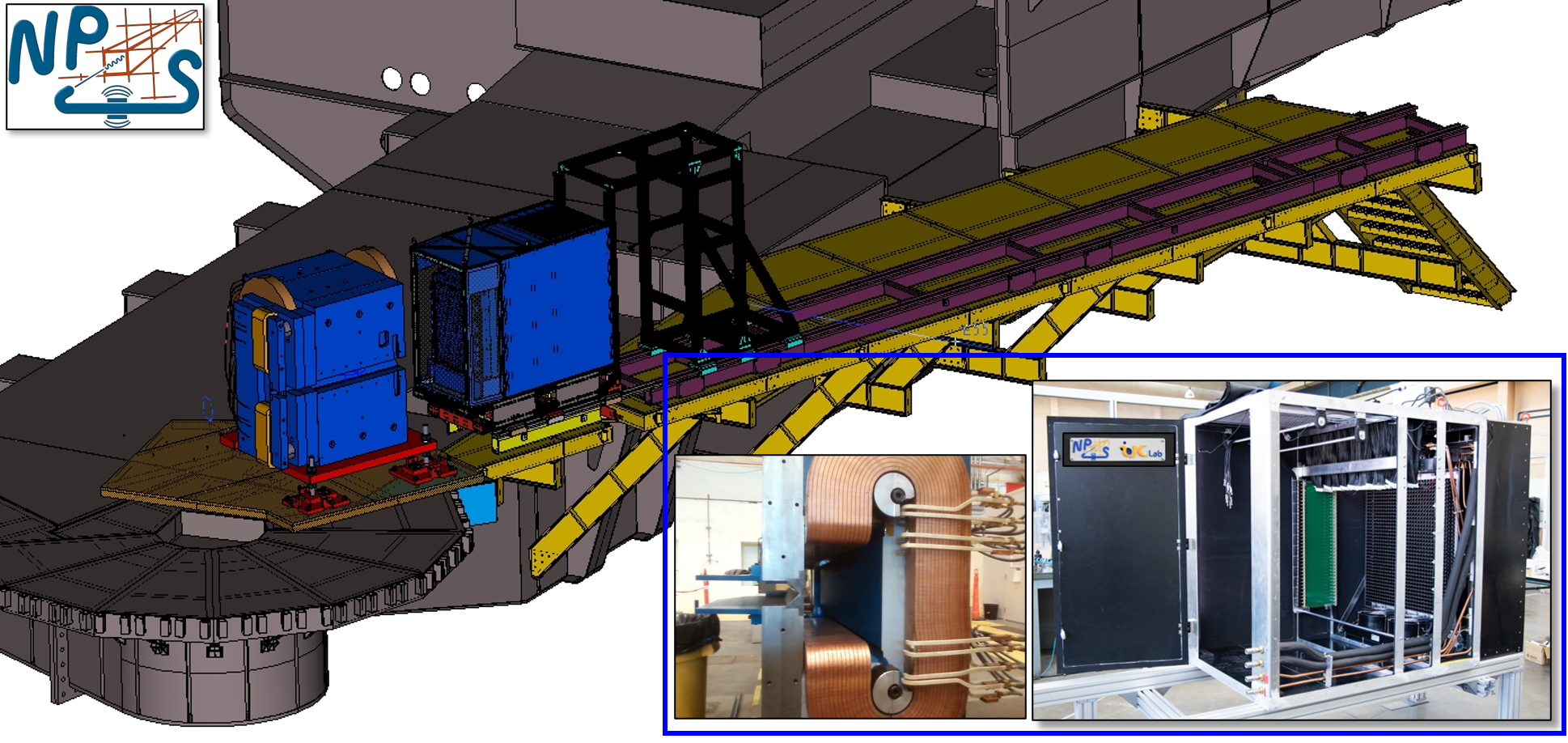}
 \caption[]{ A CAD rendition of the Neutral Particle Spectrometer (NPS). Scattered particles enter from the left. A sweeping magnet is located in front of the calorimeter to help suppress background from low energy electrons. Inset photos: (left) a close-up of the sweeping magnet, and (right) the cabinet which provides temperature stabilization and access to the phototubes and the many HV, signal, and fiber optic cables.
 }
  \label{fig:nps}
 \end{figure*} 

The 12 GeV NPS science program involves neutral particle detection over an angular range between 5.5$^{\circ}$ and 60$^{\circ}$ at distances between 3 and 11 meters from the experimental target. As operated in Hall C, the NPS will normally replace one of the base focusing spectrometers.
The experiments will use a high-intensity beam of electrons with energies of 6.6-11 GeV 
and a typical luminosity of $\sim 10^{38}~\textrm{cm}^{-2}s^{-1}$, as well as a secondary beam of photons. 
These beams will be incident on either liquid cryogenic or solid polarized ammonia targets. As seen in Fig.~\ref{fig:nps}, a vertical-bend sweeping magnet with integrated field strength of 0.3-0.6 Tm will be installed in front of the calorimeter to suppress background from low-momentum charged particle tracks originating from the target. 
Exclusivity of the reaction is ensured by the missing mass technique, the resolution of which is dominated by the energy resolution of the calorimeter. The calorimeter provides a spatial resolution of 2-3 mm and energy resolution of about 2\%/$\sqrt{E}$.
The NPS consists of 1080 PbWO$_4$ crystals that form an array of $30 \times 36$ modules. The details of mechanical assembly and commissioning of the NPS can be found in Ref.~\cite{Horn:2019beh}. The radiation hardness and good optical quality of the lead tungstate crystals are both critical.  

While much can be learned about 3D hadron structure with high-intensity virtual photons (electron beams), the NPS combined with the CPS offers a complementary approach to small cross sections with high-intensity real photon beams~\cite{Day:2019qdz}. 
Deployment of the NPS at higher energies would significantly extend measurements of DVCS, DVMP and SIDIS (the latter two involving neutral meson final states at large $x$), while the combination of NPS and CPS would allow for a substantial increase in kinematic reach for polarization observable measurements in TCS, WACS, and neutral pion photoproduction. 


Yet another opportunity for the future use of NPS is the possibility of
triple-coincidence experiments along with the SHMS and HMS spectrometers. Such experiments can study hyperon self-polarization via the rarely used $\Lambda^{0} \rightarrow \pi^0 n$ decay channel, allowing unique measurements of the $A$ dependence of $\Lambda^{0}$ self-polarization and the role of strange quark spin, orbital angular momentum and final state interactions. In addition, with a focal plane polarimeter in the SHMS, the $\Lambda^{0}$ and $\Sigma^{0}$ parity-violating decay parameters  and their ratios can be measured as a probe of CP and T-reversal symmetries. These possibilities are further discussed in Section~\ref{hyperondecay}.      



\subsection{Big Bite and Super Bigbite Spectrometers}

%
%

\begin{figure}[htb!]
 \centering 		
 \includegraphics[width=0.55\columnwidth]{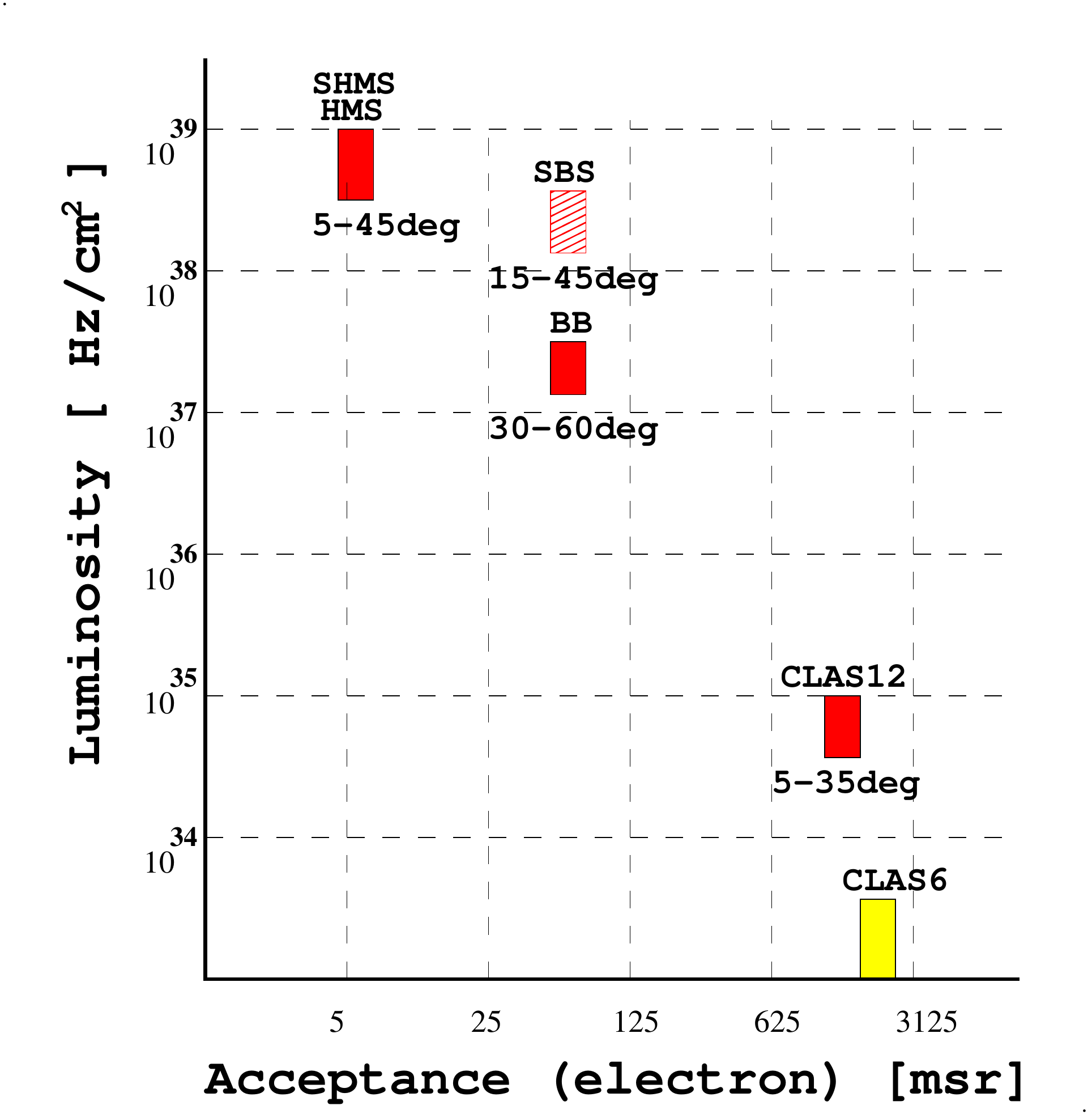}
 \caption{The maximum luminosity vs acceptance landscape for magnetic spectrometers at JLab~\cite{BW-2015}.}
\label{fig:lum_vs_acc}
\end{figure}

The BigBite (BB) and Super BigBite (SBS) Spectrometers will become available to Hall C when a series of nucleon elastic form factor experiments in Hall A is completed after 2024. 
Medium acceptance magnetic spectrometers such as these have been proven to be productive especially in experiments with polarized targets. 
As is shown in Fig.~\ref{fig:lum_vs_acc}, the SBS fills a gap between the high precision focusing spectrometers and
a large acceptance toroidal spectrometer such as CLAS12.
The use of SBS enables experiments at the full luminosity of the $^3\text{He}$ polarized target for both longitudinal and transverse directions of the polarization.
The combination of BB and SBS provides a good match for two-body final state reactions such as $e+N\rightarrow e+N$, or for pseudo two-body final states such as SIDIS in the high $z$ ($= E_{\textrm{hadron}}/\nu$) regime.

As already discussed, a number of important goals in hadron physics can only be achieved by the study of electron and photon interactions with polarized targets.
The maximum usable total luminosity for such targets is currently of order $10^{36}~\textrm{cm}^{-2}\textrm{s}^{-1}$,  
which is two orders of magnitude below that of the high power cryotargets. 
For a reaction such as elastic scattering which does not require excellent momentum resolution to isolate the final state, this allows the use of a dipole-based spectrometer with moderate solid angle and large momentum acceptance. In such a scenario, the overall experimental figure of merit can be 10 -- 20 times higher than using a small acceptance, high resolution, focusing spectrometer. 

In addition, several proposed experiments based on the Sullivan process require detection of one or more soft protons ($\sim$100 MeV/c). This requires specialized, ``thin'' recoil detectors close to the target where the high rates 
limit the maximum luminosity. Such experiments would therefore benefit from a medium acceptance spectrometer like SBS or BB to detect the electron~\cite{TDIS}.
In the following, we present the characteristics and potential applications of these two spectrometers.

\begin{table}[htb!]
\caption{Typical parameters of the BigBite Spectrometer (BB).}
\label{tab:BB}
\centering
\begin{tabular}{cc}
\hline 
Momentum        &  above 0.2 $\textrm{GeV}/c$ \\  
Momentum Resolution & typically 0.5-1\% \\
Solid angle     &  50-75 msr \\  
in-plane angle acceptance  &  $\pm 3.4$ deg \\   
out-of-plane angle acceptance  &  $\pm 11$ deg. \\  
weight          &  $\sim$23 tons   \\  
\hline

\end{tabular}
\end{table} 

\subsubsection{BigBite spectrometer (BB)}

%

The original BB spectrometer was used for detecting particles with the relatively low central momentum of 500 $\textrm{MeV}/c$ at the NIKHEF facility~\cite{DELANGE1998182, *PhysRevLett.95.172501}.
After the dipole magnet was moved to JLab, it was provided with an updated detector package and was successfully used in several experiments in 
Hall A~\cite{PhysRevLett.99.072501, PhysRevLett.105.262302,
PhysRevLett.107.072003, PhysRevLett.108.052001, PhysRevLett.113.022502, PhysRevLett.114.192503, PhysRevD.94.052003}.
Over the years, the detector package has evolved. 
The present detector package includes five layers of two-dimensional Gas Electron Multiplier (GEM) chambers, a 500 PMT gas Cherenkov counter, a two-layer segmented lead-glass based calorimeter, and a 90-paddle timing hodoscope.

The magnet has a 1 Tesla-meter field integral. With GEM tracking, the momentum resolution is typically 0.5-1\%. At large scattering angles, the magnet can be installed as close as 120 cm from the target, providing a solid angle above 50 msr. (The solid angle increases to $\sim$75 msr if the gas Cherenkov counter is not required.) These and other parameters of the BB are presented in Table~\ref{tab:BB}.

\subsubsection{Super Bigbite Spectrometer (SBS)}

\begin{figure*}[htb!]
 \centering 
 \includegraphics[width=0.75\columnwidth]{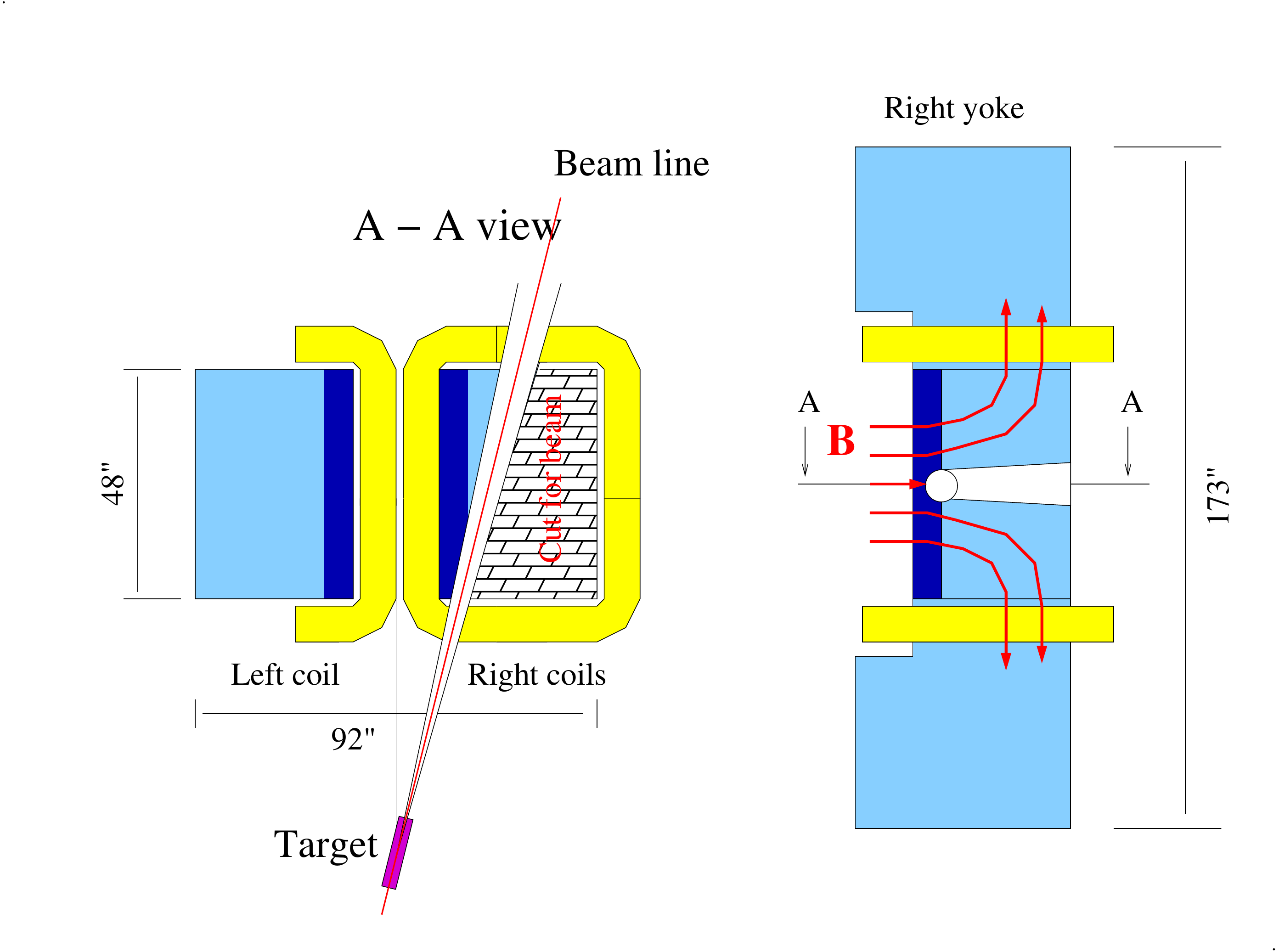}
\caption{The concept of the SBS magnet~\cite{SBS-par}. The left figure is a top view
of a horizontal cross-section through the magnet center. The right figure is a side view of the right yoke looking downstream which shows the slot which has been machined to allow the SBS to reach small angles with respect to the primary electron beamline.}
\label{fig:SBS} 
 \end{figure*}

In many potential experiments with the 12 GeV era CEBAF, a relatively large solid angle spectrometer is needed to detect secondary particles emitted at small angles $\theta$ with respect to the beam direction. 
The width of the magnet yoke typically leads to a limit on the minimal spectrometer angle, which is significant for a dipole-based, non-superconducting spectrometer. 
A new concept was suggested in the 2007 PR12-07-108 proposal to JLab PAC32.  
It took advantage of the relatively low field inside the magnet yoke and proposed to make an opening for the beamline parallel 
to the magnetic field direction. (See Fig.~\ref{fig:SBS}.)
As a result, the large dipole could be located a short distance from the target, achieving the required large solid angle at relatively small scattering angles.\footnote{Note that the maximum solid angle is limited by $2 \pi \times \sin (\theta) \times \Delta \theta$, so even 50 msr corresponds to about 20\% of the maximum possible solid angle at $\theta \sim 15$ degrees and $\Delta \theta = 8$ degrees.}



The Super Bigbite Spectrometer (SBS) is based on a large 48D48 dipole magnet donated by Brookhaven National Laboratory. (See the SBS parameters in Table~\ref{tab:SBS}.)
The mounting of this 100-ton magnet near the target is not an easy task due to the voluminous support structure required for the stability of a several meter tall device. 
This problem was solved in Hall A initially by attaching the dipole to a heavy counterweight. 
Being able to pass the beam through the right yoke, while being able to move the SBS over a range of angles, 
requires a hybrid coil system with a saddle coil on one side and two racetrack coils on the beamline side as detailed in Fig.~\ref{fig:SBS}.
 
 The field component transverse to the beam direction is relatively low and could be reduced to an acceptable level by a thin iron shield.
However, the longitudinal field component presented in the fringe areas would lead to complete saturation of the shield.
A solution was found which used a double layer shield where the outer layer is made of a set of separated iron rings,
 allowing reduction of the longitudinal field on the inner layer.
 
\begin{table}[htb!]
\caption{Typical parameters of the Super Bigbite spectrometer, Ref.~\cite{SBS-par}. (The solid angle varies with angle setting as this determines the minimum distance to the target.)}
\label{tab:SBS}
\centering
    \begin{tabular}{cc}
\hline
Momentum        & above 2 $\textrm{GeV}/c$	\\
Solid angle     &  50-75 msr        \\
in-plane angle acceptance  & $\pm 4.8$ deg.    \\
out-of-plane angle acceptance &    $\pm 12.2$ deg. \\
 weight   &  $\sim$200  tons  \\ \hline
 \end{tabular}
\end{table} 



The BB and SBS spectrometers can be used with the target at the nominal Hall C pivot position as well as at a downstream position.
The BB magnet was already used in Hall C near the nominal pivot during the Wide Angle Compton scattering experiment~\cite{PhysRevLett.115.152001}. 
However, installation at a position downstream of the pivot allows access to a wider range of scattering angles.


	

\subsection{HCAL and ECAL}


%

Experiments with exclusive reactions simultaneously at large $s$, large $-t$, and large $u$ require luminosity of order $10^{38}~\textrm{cm}^{-2}\textrm{s}^{-1}$. This can be realized with sufficiently segmented detectors and a high energy trigger threshold, the latter being 
a natural feature of calorimeters.
A hadron calorimeter (HCAL) was prepared for high momentum transfer form factor experiments in Hall A for use with the SBS.
An electromagnetic calorimeter (ECAL) is under construction for the proton GEp experiment (E12-07-109).
The large size of these calorimeters could make  them attractive for other programs.
For example, the TDIS experiment plans to use HCAL for precision determination of the vertex detector efficiency.
Also, the ECAL and HCAL modules could be used in the Coincidence Parity experiment discussed in 
section~\ref{sec:COP}.
We present here the basic characteristics and potential applications of these two calorimeters in experiments in Hall C. 


\subsubsection{Hadron Calorimeter (HCAL)}

In neutron form factor experiments, recoil neutrons have often been detected by layers of $\sim$10 cm thick scintillator counters. These allow good time resolution (0.25~ns), but even with a low energy threshold have limited efficiency per layer due to the nearly 80 cm nuclear interaction length 
~\cite{PhysRevLett.92.042301, PhysRevC.73.025205,Glazier-2005,PhysRevLett.105.262302,PhysRevLett.111.132504}.
By contrast, the CLAS experiment~\cite{PhysRevLett.102.192001} used a Pb-scintillator electromagnetic calorimeter to achieve neutron detection efficiency up to 60\%~\cite{AMARIAN2001239}. 
For kinetic energies above $\sim$1 GeV, a sufficiently thick hadron calorimeter can provide efficiency close to 100\% for protons and neutrons.
High efficiency provides not only smaller statistical errors but also reduces the systematic uncertainty in the ratio of the neutron to proton detection efficiencies which is critical for the important class of neutron form factor experiments based on $d(e,e^{\prime}n)/d(e,e^{\prime}p)$.

\begin{figure*}[htb]
 \centering 
 \centering 		
 \centering 		
\includegraphics[width=0.85\columnwidth, trim = {25mm 30mm 0mm 20mm}, clip ]{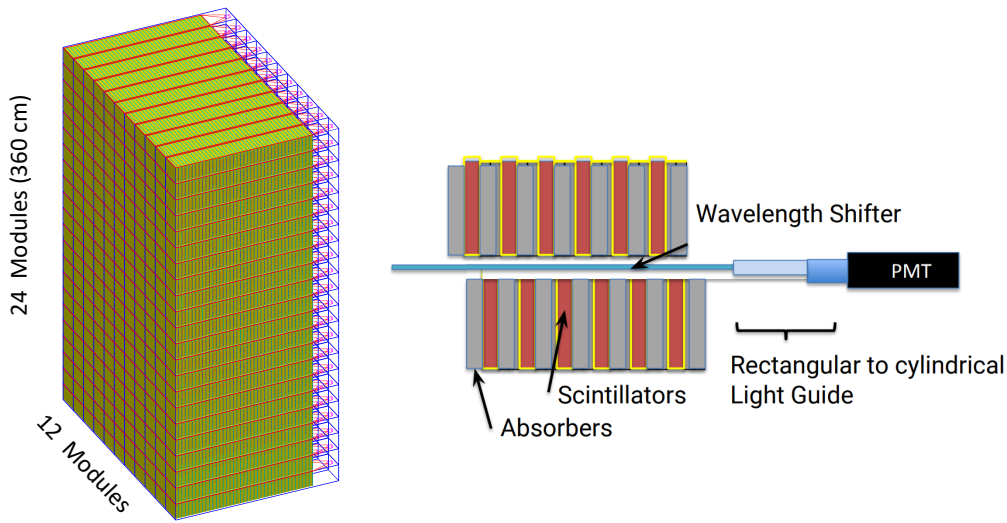}
 \caption{
 (Left panel) Schematic of the 288-block assembly, and (right panel) the method of light collection in an individual HCAL module.}
 \label{fig:HCAL} 
\end{figure*}


The hadron calorimeter used in SBS is HCAL. It is a sampling calorimeter with a wave length shifter for light collection.
HCAL includes 288 modules in an array of 12x24 with 15.5 cm by 15.5 cm cross sections of each~\cite{franklin_hcalj_2014}.
The total active area of the HCAL is about 6.5 m$^2$.
The geometry of an HCAL module is shown in Fig.~\ref{fig:HCAL}.
The concept of the calorimeter is similar to that of the COMPASS hadron detector~\cite{calo-compass} but with a few design changes: fast, larger diameter PMTs, faster scintillator and wave length shifter, and a novel geometry light guide.
The HCAL detector provides high efficiency 
and a good time resolution of 0.5-1~ns. The basic parameters of HCAL are summarized in Table~\ref{tab:HCAL}.

\begin{table}[htb]
	\caption{Components of the hadron calorimeter.}
	\centering
   	 \begin{tabular}{cc} 
\hline
Number of Modules    & 288 \\
                     &  (each is a 12x24 array)                       \\
 Module area         & 15.5x15.5 cm$^2$                       \\  
Layers of Fe-Scin   & 40                                   \\  
Steel per layer      &  two 1.27x7.0x14.9 cm$^3$      \\  
Scintillator per layer  &  two 1x6.9x14.8  cm$^3$       \\
                     &  FNAL 0.5\% POPOP                       \\  
WLS per module       &  0.5x13.6x94.8 cm$^3$        \\
                     &   BC-484                                \\
PMT type            &  XP2262 and XP2282                     \\
\hline
 	\end{tabular}
	\label{tab:HCAL}
\end{table}


\subsubsection{Electromagnetic Calorimeter (ECAL) }

In several JLab experiments, lead-glass based calorimeters have been used, see e.g. Refs.~\cite{PhysRevLett.98.152001, PhysRevLett.104.242301, PhysRevLett.115.152001}.
The largest one was constructed in Hall D for the GlueX experiment~\cite{MORIYA201360}.
Several types of lead-glass were used in these calorimeters : TF1, TF101, and F8.
These types of lead-glass have good light transparency but suffer from a significant loss of transparency when the radiation dose exceeds only 1-10~kRad~\cite{SCHAEFER2012111, *BALATZ2005114, *Inyakin:1981tg}.
The latter characteristic normally curtails the use of lead-glass detectors at high luminosity. 
Radiation damage induced by electromagnetic radiation can be cured by exposure of the lead-glass to UV-light and/or elevated temperature, 
see Refs.~\cite{HAMILTON201117, MORIYA201360}.
However, both methods require interruption of data taking for a significant time for the dismounting of the lead-glass blocks, curing and reassembly.

In 2010, the concept was presented for the ECAL: a lead-glass calorimeter to be operated at constantly elevated temperature~\cite{lead-glass-highT-2013}.
A detailed study showed that 225$^\circ$C is a sufficiently high temperature to keep the lead-glass transparent at the dose rate expected for the SBS/GEP experiment~\cite{lead-glass-study-2014}, while being too low a temperature for black body radiation to contribute to the PMT dark current.
 
Because an ordinary PMT photocathode cannot operate at a temperature above 50$^\circ$C,
a 15 cm long light guide and a cooling air jet were added between the lead-glass block and PMT.
A beam test was performed in Hall A using tagged electrons from the $\textrm{H}(e,p)e^{\prime}$ process~\cite{lead-glass-highT-2015}.
The observed energy resolution was 10\% for 1.6 GeV electrons.\footnote{The elevated temperature leads to some loss of the lead-glass transparency.}

\begin{figure}[htb]
 \centering 
 \centering 
 
  
 \centering 	
   \includegraphics[width=0.8\columnwidth, trim = 20mm 10mm 20mm 10mm]{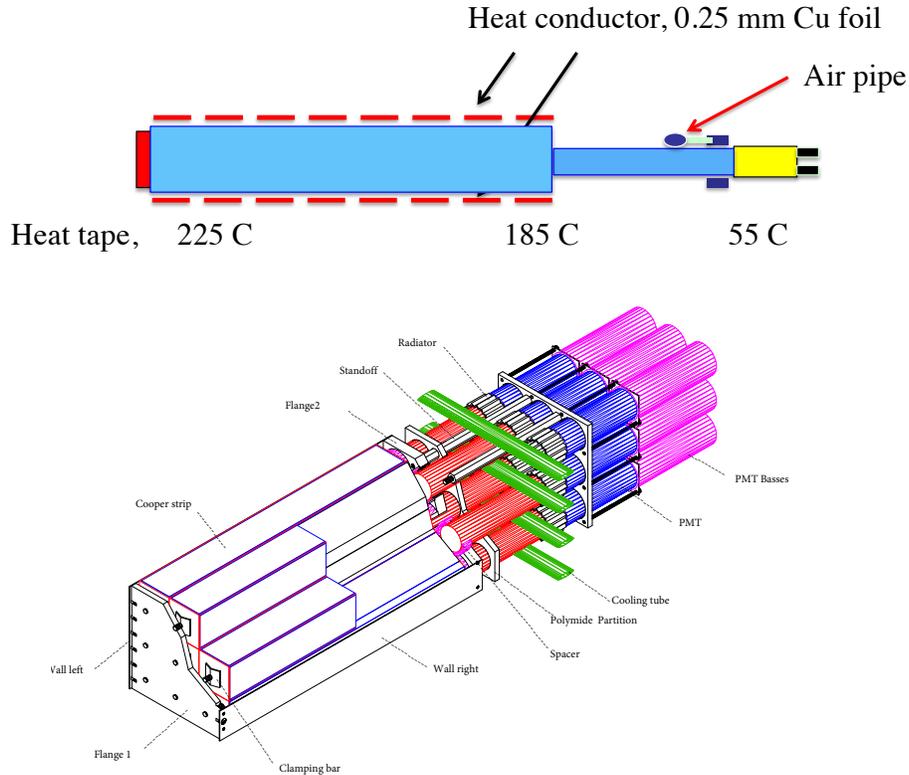}	
  \caption{
  Top - An individual heated block in ECAL.
  Bottom - A module of $3\times 3$ blocks. }
  \label{fig:ECAL}  
\end{figure}

Currently, the full-scale calorimeter is under construction by the collaboration of NCCU, JMU, YerPHI, and JLab.
The distribution of blocks in the calorimeter somewhat resembles the letter ``C'' to match the electron distribution for fixed $Q^2$ in the $G_E^p$ experiment.
The lead-glass blocks' design is shown in Fig.~\ref{fig:ECAL}. The light guide is made of BC7 glass to match the thermal expansion of the lead-glass.
Due to the high temperature, reflective wrapping of the glass employs Al foil rather than the usual aluminized mylar.
An additional layer of Cu foil is needed to increase heat conductivity along the lead-glass module.
Each group of nine blocks is assembled into a super-module.
The frame of the super-module with Ti side walls allows attachment of the PMTs to the light guides
and optically transparent soft plastic pads and mounting of the PMTs.
The total active area of the ECAL detector is 3.2 m$^2$.

 \subsection{A compact solenoidal spectrometer for deep exclusive measurements}
 \label{sec:css}
As pointed out earlier in this document, Hall C has a unique role to play in high luminosity, deep exclusive measurements. While HMS and SHMS will be capable of making high precision measurements over limited regions of phase space, harnessing the full potential of JLab deep exclusive measurements requires a 2$\pi$ acceptance device capable of operating at the highest luminosities available in Hall C. Furthermore, exclusive reactions require very good vertex resolution for establishing exclusivity. 

\begin{figure*}[hbt!]
  \centering
\includegraphics[width=0.7\textwidth]{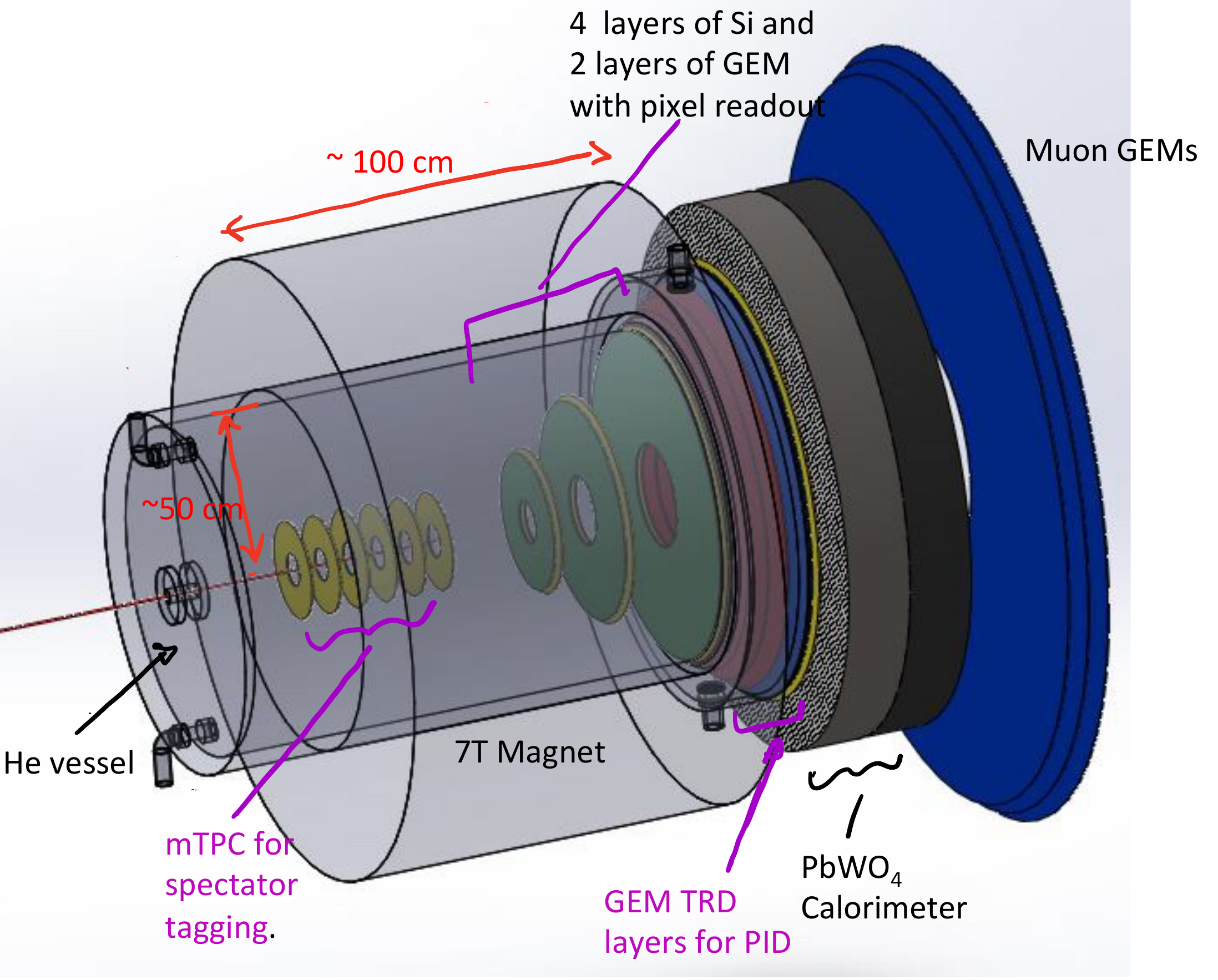}
 \caption[]{A CAD rendition of the proposed compact solenoid spectrometer. The bore of the magnet will be filled with helium gas at STP to reduce multiple scattering. The target will be located at the center of the mTPC. The calorimeter and the muon chambers could be moved further downstream to reduce pile-up in the calorimeter as needed.  
 }
  \label{fig:compact_solenoid}
 \end{figure*}

The Compact Solenoid Spectrometer proposed here has been optimized to meet the above requirements. 
The CAD model for the proposed device is shown in Fig.~\ref{fig:compact_solenoid}. This spectrometer design is based on a compact, 7 T solenoidal magnet with an approximate length in the range of 100-150 cm and a radius in the range of 50-75 cm. The cost of a super-conducting high field magnet of this size is about \$5 million. The relatively short length of this magnet is compensated by the high field strength. In fact, the $\int B \, dl$ range for this magnet is higher than that of much larger solenoidal magnets such as the BaBar and CLEO magnets. The small size of the magnet, and hence the short path length from the target to tracking detectors, provides very good vertex resolution. The vertex resolution will be further improved by filling the small magnet bore with He. Simulations show that with these factors combined, this proposed device will have a vertex $z$ resolution of about 250 $\mu$m, and a momentum resolution of about 0.3\%.

The relatively small size of the magnet permits instrumenting the spectrometer with state of the art detectors that are highly cost effective. Thus, the area that needs to be covered by detectors is only a small fraction of the area of a traditional large acceptance spectrometer. Given that the detectors are close to the target, the rates at the detectors are expected to be high. Simulations indicated that for a luminosity of $10^{38}~\textrm{cm}^{-2}\textrm{s}^{-1}$, the rates at the first detector wheel could be up to 10 MHz/cm$^2$ which is about a factor of 20 higher than the rates expected in the SoLID spectrometer. However, using state-of-the-art Silicon and GEM  detectors with pixel readout, this high rate could be managed. The suite of detectors for this spectrometer includes (i) a multi-TPC with 6 pixel GEM based readout disks for spectator tagging, (ii) four layers of Silicon (Monolithic Active Pixel Sensor -MAPS) detectors which have only 0.05$\%$ radiation length and a  fine pixel pitch of 10 $\mu$m, (iii) two layers of GEM detectors with pixel readout, 
(iv) GEM based Transition Radiation Detector (TRD) for $e-\pi$ particle ID, (v) a lead tungstate calorimeter, and (vi) two layers of GEM detectors behind the calorimeter for muon detection. 
 
\section{A Sampling of Potential Hall C Experiments}
\label{phys_prog}

In this section, we outline some potential large experimental programs that could be carried out in Hall C. 

\subsection{ Precision Measurements in Exclusive and Semi-Inclusive Scattering}\label{sect:precLT}

Hall C has a uniquely important role to play in the EIC era, particularly in the realm of precision L-T separation measurements.  Conventional Rosenbluth separations are impractical at the EIC.  This is because systematic uncertainties in $\sigma_L$ are magnified by $1/\Delta\epsilon$, where $\Delta\epsilon$ is the difference in the virtual photon polarization parameters at high and low center of mass energies.  To keep the systematic uncertainties in $\sigma_L$ to an acceptable level, $\Delta\epsilon>0.2$ is arguably required to keep the error magnification factor below 5.  However, $\epsilon<0.8$ can only be accessed with very low proton ring energies (5-15 GeV), where the luminosities will be too small for a practical measurement.  Since it will not be practical for EIC to do L-T separations, physicists will need to rely on an extrapolation of L-T separated data from Hall C.  Thus, high quality data from Hall C will be essential for the interpretation of the EIC exclusive and semi-inclusive programs.

\subsubsection{Exclusive Scattering}

The currently-approved program of Hall C exclusive measurements will benefit from a modest upgrade of the electron beam energy.  As one example, E12-19-006 is a measurement of the pion form factor to $Q^2$=6.0 GeV$^2$ with high precision, and to 8.5 GeV$^2$ with somewhat larger experimental and theoretical uncertainties.  The $Q^2$=8.5 GeV$^2$ measurement in particular suffers from the limited $\epsilon$ lever arm allowed by a maximum 11~GeV beam energy, and would benefit greatly from the possibility of extending this measurement with higher beam energy using the SHMS+HMS.  Deep Exclusive Meson Production (DEMP) scaling tests are another example of L-T separations that would benefit from the improved lever arm enabled by a beam energy upgrade.

\begin{figure}[hbt]
  \centering
\includegraphics[width=0.7\textwidth]{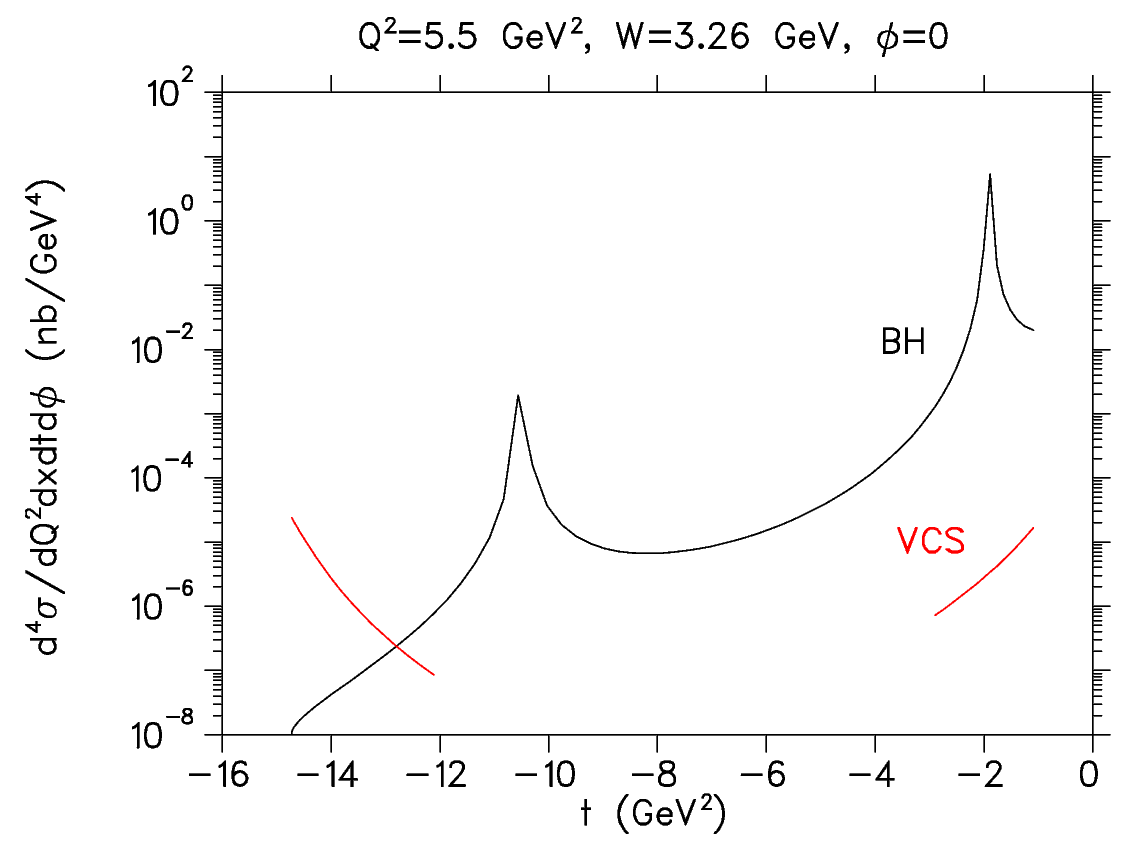}
 \caption{$t$ evolution of Virtual Compton Scattering (VCS) and Bethe-Heitler (BH) processes at $Q^2=5.5$ GeV$^2$, $W$=3.26 GeV.  BH dominates at small $-t$, and is suppressed by $\sim 10^6$ at the extremely large $-t$ (backward photon angles) accessible in $u$-channel measurements.}
 \label{fig:u-dvcs}
 \end{figure}

A new direction for the Hall C exclusive program is the study of $u$-channel reactions which allow access to the valence+sea component of the nucleon wavefunction in a particularly clean manner. The long-term goal for this program is the construction of dedicated detectors in the backward-angle kinematics region ($90<\theta_{Lab}< 180^\circ$). This would allow measurements to access backward-angle cross-section peaks in a variety of exclusive channels over a wide range of kinematics (where the missing mass technique cannot be used due to poor resolution). These results will test the kinematic onset of the backward-angle colinear factorization theorem (analogous to forward-angle GPD factorization) \cite{pire2021transition}.  The first objective depends significantly on the capacities of the HMS and SHMS to identify the $u$-channel recoil meson through missing mass reconstruction with moderate momentum.  The second objective depends critically on Hall C’s L-T separation capability, as the onset of backward-angle factorization requires the demonstration of $\sigma_T\gg\sigma_L$ as well as the requisite $1/Q^n$-scaling of $\sigma_T$ and $\sigma_L$.  Backward-angle DVCS is a particularly interesting reaction to study, as the competing Bethe-Heitler process that is dominant at forward-angles is highly suppressed at backward-angles as shown in Fig.~\ref{fig:u-dvcs}, allowing clean access to $u$-channel DVCS.  Such studies would require the HMS+SHMS in coincidence with a large solid angle calorimeter such as NPS or ECAL.

\subsubsection{Semi-Inclusive Scattering}
Little is known of the longitudinal and transverse components in the SIDIS process. To date it has been assumed that the value determined from inclusive DIS can be used, irrespective of possible dependences on quark flavor, hadron type, hadron fractional energy $z$ or transverse momentum $p_T$.  Longitudinal and transverse effects must be studied as a check of our basic understanding of the photon-bound quark reaction, especially as it may affect our understanding of transverse momentum dependent structure and fragmentation functions. An approved Hall C proposal will make the first systematic exploration of the L and T dependences of SIDIS; further studies will rely on these first experimental findings.

More generally, the next major epoch of our study of quark-gluon dynamics in SIDIS will likely focus on the ``fine structure’’ of the physics resulting from the mass differences of the up and down quarks, in other words, flavor structure. Experimental study of the nucleon structure with SIDIS continues to be plagued with being required to make assumptions about quark flavor dependence or independence in structure and reactions, even though it is known that non-trivial flavor dependence exists. These effects are very small, in some cases on the scale of the mass differences between up and down quarks (relative to the mass of the nucleon). Hence, experimental study will require high intensity beams on hydrogen and deuterium targets to precisely determine the differences in up and down quark momentum dependent distributions (transverse and longitudinal). The ability to rapidly switch between these targets provides a powerful method to reduce systematic uncertainties and obtain results relatively quickly. Detection of charged and neutral pions will be critical to understanding the flavor dependence in the momentum dependent fragmentation functions necessary for unraveling the nucleon structure. As mentioned above, an increase in beam energy will allow the study of these details over an extended range of quark and hadron kinematics, typically limited by the desire to keep the un-detected hadronic mass above that of nucleon resonances.

Precisely measured kaon SIDIS from hydrogen and deuterium provides exciting access to strange quark distributions in the nucleon and helps in the understanding of un-favored fragmentation. Here again, Hall C has a unique role due to its identical acceptance for charged pions and kaons, while such acceptances must necessarily differ in the solenoidal field planned for the EIC. A high reach in $p_T$ requires Hall C’s high luminosity, and such studies are not possible with an open geometry detector.

%
\subsection{Color Transparency}
Quantum Chromodynamics (QCD) predicts 
that hadrons produced in exclusive processes at sufficiently high 
4-momentum transfer will experience suppressed final (initial) state interactions resulting in a significant enhancement in the nuclear transparency~\cite{CT, CT_brodsky}. 
This unique prediction of QCD is named color transparency (CT), and the onset of CT is thus a signature of QCD degrees of freedom in nuclei which can help identify the transition between the nucleon-meson and the quark-gluon descriptions of the strong force. However, the energy regime for the onset of CT is not precisely known and must be determined experimentally.
The suppression of further interactions with the nuclear medium is well established at high energies as it is a fundamental assumption necessary to account for Bjorken scaling in deep-inelastic scattering at small $x_B$~\cite{FS88}. 
Moreover, the onset of CT is of specific interest as it can help identify the relevant space-like 4-momentum transfer squared ($Q^2$) where factorization theorems are applicable~\cite{strikman} thus enabling the extraction of Generalized Parton Distributions (GPDs)~\cite{Ji:1997, Radyushkin:1997}.

While empirical evidence ~\cite{clasie07, xqian10, elfassi12} conclusively confirms the onset of CT in mesons at momentum scales corresponding to $Q^2 \approx~1\,(\textrm{GeV}/c)^2$, the onset of CT in baryons remains highly sought after but as yet unobserved. In fact, the result from a recent high precision $A(e,e'p)$ experiment in Hall C~\cite{Bhetuwal:2021} shows a puzzling lack of color transparency in protons even for $Q^2$ above 10 (GeV/c)$^2$. These results probe down to a transverse size as small as $\approx$~0.05 fm in the three-quark nucleon system, placing very strict constraints on the onset of color transparency at intermediate energies and on all current models. It is critical to understand the origins of the apparent reaction dependence of this fundamental prediction of QCD, and/or the differences between three-quark and quark-antiquark states.

Most CT searches tend to choose kinematics where final state interactions are low and the quasi-elastic rates are sufficiently high. Given the apparent lack of CT in these kinematics, it is highly desirable to probe other regions of the phase space that move away from these "clean'' kinematics. It is also important to use reactions that are precisely described by nuclear physics calculations. One such reaction is $d(e,e'p)$, which is well-described over a very large phase space within the generalized eikonal approximation (GEA)~\cite{Sargsian97}. Recent results on $d(e,e'p)$ have clearly shown that final state interactions (FSI) strongly dominate the yields at large missing momentum and backward recoil neutron angles (such as $P_{r} \sim$ 400 MeV/c, $\theta_{nr}\sim$ 70$^{\circ}$ )~\cite{Yero20,Boeglin11}. These results are reproduced remarkably well by the predictions of GEA calculations. 

\begin{figure*}[hbt!]
  \centering
\includegraphics[width=0.75\textwidth]{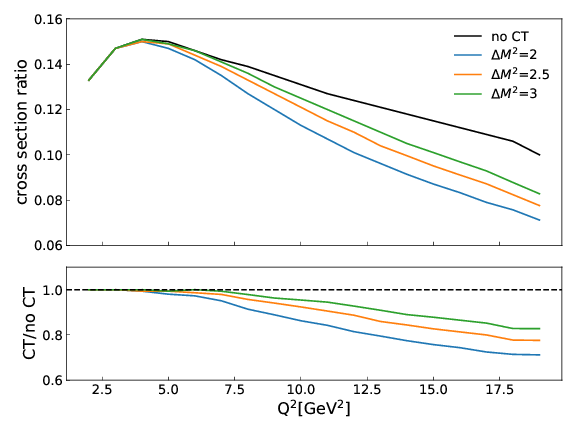}
\caption[]{(Top panel) The ratio of  $d(e,e'p)$ cross sections at kinematics dominated by FSI to those with reduced FSI for various hypotheses of the $\Delta_M^2$ parameter of the quantum diffusion model (QDF)~\cite{PhysRevLett.61.686}~(the mass square difference between the proton and its intermediate state). The three curves show the range of values not ruled out by the recent Hall C results. The "no CT" prediction shows a calculation in the GEA framework~\cite{Sargsian97} while the predictions that include CT~\cite{Sargsian03} are based on the QDF. (Bottom panel) The cross section ratios from the top panel are shown as ratios relative to the "no CT" prediction.}
 
  \label{fig:deepct}
 \end{figure*}

In this regime of large FSI, there exists an interference between the Born term and the rescattering amplitude of the cross section. The calculation of the cross section shows a significant increase when the rescattering is included. (The square of the rescattering amplitude is dubbed the "double scattering" term~\cite{Frankfurt:1994kt}.) These rescattering effects would decrease in the presence of a point-like configuration (PLC). 
On the other hand, the FSI are found to be suppressed at small missing momentum (such as $P_{r} \sim$ 200 MeV/c). Therefore, the ratio of $d(e,e'p)$ yields at large missing momentum to yields at small missing momentum, where both are measured at backward recoil neutron angles $\theta_{nr}\sim$ 70$^{\circ}$, is an ideal probe for the onset of CT. This ratio as a function of $Q^2$ would be very sensitive to the reduction in FSI predicted by CT as indicated by the calculation shown in Fig.~\ref{fig:deepct} and can be measured with the standard Hall C equipment. Given the small rates at these extreme kinematics characterized by double scattering, and the need for high resolution and high precision, such measurements can only be carried out in Hall C.

Another sensitive probe of CT is the asymmetry $A_{LT}$ of the $A(e,e'p)$ yield for proton angles to the left and right of the three-momentum transfer ($\vec{q}$) direction, measured at high missing momentum. The observable $A_{LT}$ is very sensitive to FSI and thus when measured as a function of $Q^2$ would be another important probe of the reduction in FSI predicted by CT. Finally, the normal component   of the recoil proton polarization ($P_n$) in $A(\vec{e},e^{\prime} \vec{p})$ is essentially an FSI meter. In the absence of any nuclear medium effect, $P_n$ should be zero, thus $P_n$ is another probe of CT. Using a focal plane polarimeter (FPP) in the hadron arm spectrometer, one can search for the onset of CT in the  $Q^2$ range of 1 - 10 GeV$^2$. For these measurements, $p(\vec{e},e^{\prime} \vec{p})$ can be used for self-calibration and false asymmetry measurements~\cite{arun91,anklin94}. It is estimated that a 10\% measurement on a range of nuclear targets can be completed in about 500 hours of beam time over a $Q^2$ range of 1 - 10 GeV$^2$.

Measuring the ratio of large to small missing momentum cross sections in $d(e,e'p)n$, the asymmetry $A_{LT}$, and the recoil polarization $P_n$, an extensive $A(e,e'p)$ program seeking FSI variation vs $Q^2$ in non-traditional kinematics is envisioned as a strong potential Hall C program.
 
\subsection{Lambda Hypernuclear Spectroscopy}

Recent astronomical observations of gravitational
waves and X-ray pulsars have provided new constraints on the mass and radius of neutron stars, which consist of the most dense material in the
universe. Our knowledge of baryonic forces, which have been
microscopically studied by various nuclear experiments, is now being challenged by such macroscopic observations. One example is
that conventional baryonic potential models with hyperons cannot
explain the existence of heavy (two solar mass) neutron stars. Precise spectroscopy of hypernuclei is one of key studies to provide
information on the three-body baryon repulsive force which could make
neutron stars' equation of state stiff enough to support such high masses without collapse into a black hole. 
Currently, electroproduction of $\Lambda$ hypernuclei is the only 
technique which can provide the absolute binding energies of hypernuclei
with sub-MeV resolution over a wide range of nuclear masses. The high quality electron
beam of CEBAF has enabled us to perform such experiments~\cite{Nue, Tang}.

Hypernuclei are produced and investigated by the $(e,e^{\prime}K^{+})$
reaction at JLab~\cite{GOG1}.  A magnetic spectrometer which has a
short optical length is necessary to measure the $K^{+}$ efficiently due to its decay lifetime of only $c\tau=3.71$~m.  The high
resolution kaon spectrometer (HKS) was designed and fabricated to
detect $K^{+}$ from hypernuclear production with a momentum
resolution of $\Delta p/p =2 \times 10^{-4}$ (FWHM)~\cite{GOG2,Fujii}.
Hypernuclear experiments with HKS successfully provided precise data
on hypernuclear structure.  These data proved that Charge Symmetry Breaking (CSB) in the p-shell mass
region is small, as theories predicted~\cite{Gogami:2021}.  It was found that
the data on hypernuclear structure tell us
interesting features about the nuclear core with which a $\Lambda$ is weakly
coupled. 
For example, we observed that the unstable state of the nuclear core becomes stable against the strong-interaction decay by 
neutron emission due to a glue-like role of $\Lambda$ in the
hypernucleus $^{7}_{\Lambda}$He~\cite{GOG4}.  Also, the measurement of $^{9}_{\Lambda}$Li indicated a development of the
cluster structure for the particular state of the nuclear core thanks
to the behavior of $\Lambda$ as a probe~\cite{GOG5}.  The data on  $^{10}_{\Lambda}$Be led to the need for a model-space
extension of the shell-model calculation to reproduce characteristic
level structures due to the presence of a $\Lambda$ in the
nucleus~\cite{GOG6,UMEYA}.  We aim to improve the technique to
achieve better accuracy and precision for further studies at
JLab.

\begin{figure*}[!htbp]
  \centering
  \includegraphics[width=10cm]{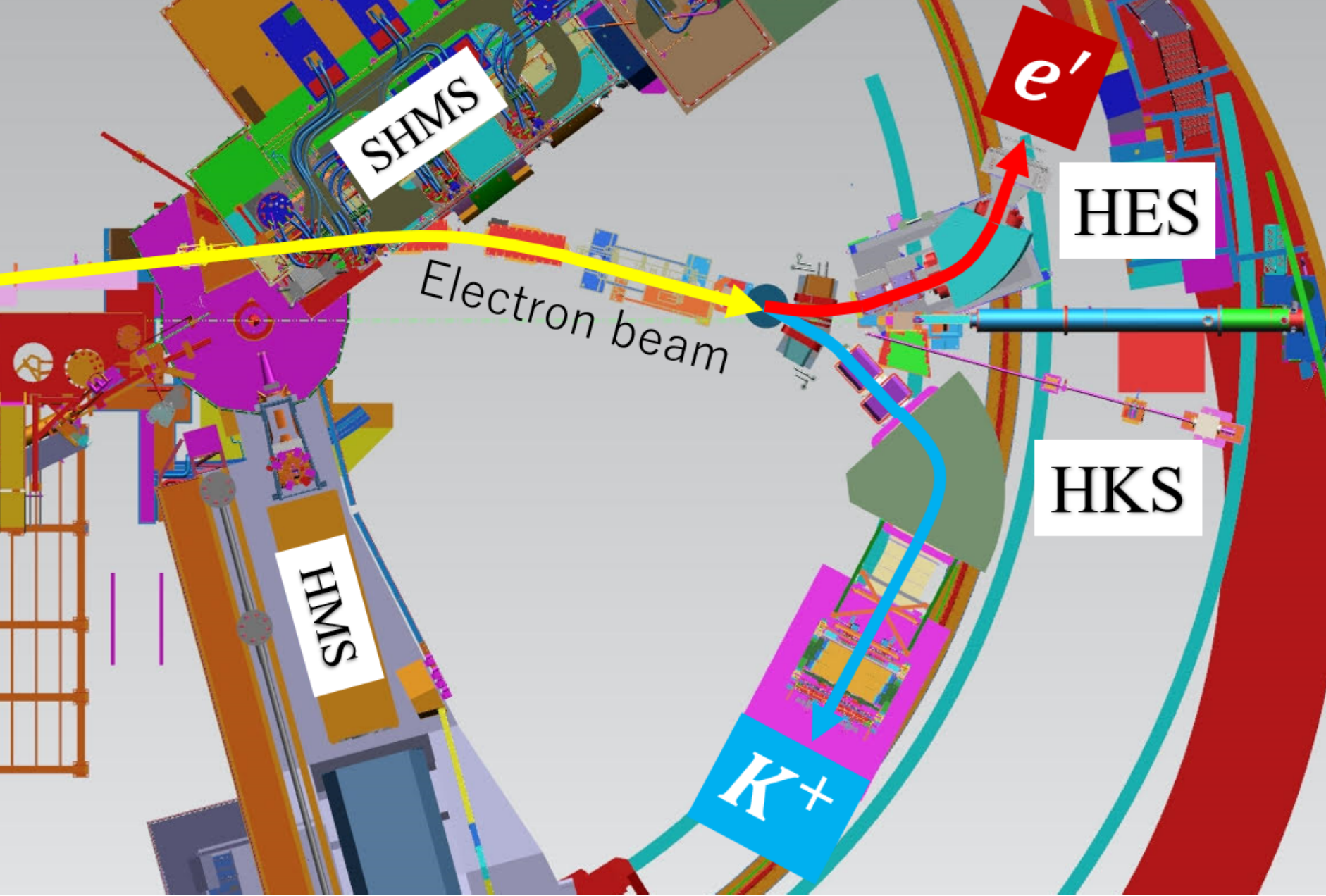}
  \caption{Schematic of one concept for the experimental setup for hypernuclear missing-mass spectroscopy Hall C. 
    A scattered electron and $K^{+}$ from the $(e,e^{\prime}K^{+})$ reaction
    are measured in HES and HKS, respectively.  
    HES is bending horizontally in the figure, however  
    the vertical bending option is being considered for extended gas targets.}
  \label{fig:hyper_setup}
\end{figure*}

There may be several options for a spectrometer setup to perform
hypernuclear experiments in Hall C, but only two options are presented 
here.  The first option is the use of HES and HKS for $e^{\prime}$ and
$K^{+}$ detection, respectively. (See Fig.~\ref{fig:hyper_setup}.) 
However, in order to use extended gaseous targets, the HES may need to be modified
to become a vertical bending spectrometer rather than a horizontal bending
spectrometer.
Data analyses for the gas targets require a
reasonably good resolution of the production position along the beam
direction which is reconstructed by the transfer matrix and is used
for momentum and angle reconstructions.\footnote{For instance, HRS, a vertical bending spectrometer in Hall A, has sufficient capability
for this kind of analysis~\cite{GOG3,SUZUKI,PAND}. But Hall A is likely to be unavailable due to the MOLLER experiment and the SOLID program.}  The second option is to
use SHMS and HKS.  In this case, the SHMS is already a vertical
bending spectrometer.
However, SHMS would limit the resulting missing-mass resolution due to
its limited momentum resolution of order $\Delta p/p =
10^{-3}$ (FWHM).  Monte Carlo simulations are ongoing to find good
options taking into account the expected missing mass resolution,
yield, and signal-to-noise ratio assuming a reasonable beamtime.

The systematic extension of such JLab studies to a variety of hypernuclear species is promising and will provide unique data which cannot be
investigated at other facilities.  One of the important studies is the
energy level comparison among hypernuclei which have a large
difference in neutron numbers to investigate the isospin dependence of
the $\Lambda$NN interaction.  The first attempt to investigate the
isospin dependence of the $\Lambda$NN interaction is being
prepared as E12-15-008.  Spectroscopy for other
hypernuclear systems from the stable isotopes of Ni, Zr, Sn, Sm, etc. could be the next challenges.  In summary, such a Hall C program could provide important and unique
data for the study of the baryonic force which includes the strangeness
degree of freedom in the SU(3) flavor symmetry.

\subsection{Precision Measurements using Polarized Targets }\label{poltarphysics}

\subsubsection{Solid polarized $NH_3$ and $ND_3$ targets}
 

The spin-dependent structure functions $g_1$ and $g_2$ are typically extracted from doubly polarized inclusive scattering.  Historically, measurements of the spin structure function $g_1$ 
at large momentum transfer $Q^2$ have 
provided direct tests of QCD via the Bjorken sum rule, and also revealed that only a small fraction of the nucleon spin is carried by the valence quarks via the Ellis-Jaffe Sum Rule.

The spin structure function $g_2$ is generally not well determined for either the proton or neutron.  This is in contrast to all other structure functions, and
arises primarily from the technical challenge of propagating an electron beam through a transverse magnetic field.
The relatively poor knowledge of $g_2^p$
is particularly unsatisfying.   Tests of the Burkhardt-Cottingham (BC) sum rule~\cite{Burkhardt:1970ti}:
\begin{eqnarray}
\Gamma_2(Q^2) \equiv \int_{0}^{1} g_2(x,Q^2) = 0
\label{eq:BCSUM}
\end{eqnarray}
have provided mixed results, and high precision   data at large Q$^2$ is needed.

The twist-3 matrix element $d_2$ measures the deviation of the $g_2$ structure function from leading twist behavior:
\begin{eqnarray}
d_2(Q^2) = 3\int_0^1 x^2\left[g_2(x,Q^2) - g_2^\textrm{WW}(x,Q^2)\right] dx
\end{eqnarray}
As such, it is
sensitive to quark-gluon interactions beyond the simple quark parton model.
Experimentally, the $x$-weighting ensures
that by far the most significant contribution to the integral comes from
the large $x$ region that is kinematically accessible at JLab. It reduces to the twist-3 matrix element at large $Q^2$.

      The nucleon polarizabilities at $Q^2=0$ represent the response of the nucleon to an external electric and magnetic field. The forward spin polarizabilities $\gamma_0$ and $\delta_{LT}$ generalize this to finite $Q^2$ where we are instead interested in the interference of the electric and magnetic fields induced by the virtual photon.  
      The doubly-virtual Compton scattering dispersion relations are used to form a low-energy expansion of the spin-flip Compton amplitudes, giving rise to a number of spin structure function moments. The next-to-leading order term of the low energy expansion contains the generalized longitudinal-transverse (LT) spin polarizability:
\begin{align}
\label{DLTEQ}
\delta_{LT} (Q^2) = \frac{16\alpha M^2}{Q^6}\int_{0}^{x_0} x^2{\bigg (} g_1(x,Q^2) + g_2(x,Q^2){\bigg )} dx
\end{align}
      These polarizabilities are extracted from linear combinations of the spin structure functions weighted by $x^2$ and integrated over the entire kinematic range. 
      It is expected that at large $Q^2$, the spin polarizabilities become independent of $Q^2$, and the DIS Wandzura-Wilczek relation leads to a relation between $\gamma_0$ and $\delta_{LT}$\cite{Drechsel:2002ar}  : $\delta_{LT}(Q^2) \to \frac{1}{3} \gamma_0(Q^2)$.  This is as-yet untested at large $Q^2$ for either a proton or neutron target.
      
      

    The newly developed tensor polarized target discussed in Section \ref{sect:poltarg} will be used during the JLab 12~GeV era to measure the leading twist deuteron structure function $b_1$, the quasi-elastic tensor asymmetry $A_{zz}$, and the elastic tensor analyzing power $T_{20}$. These observables will shed new light on fundamental questions of hadronic matter by probing the tensor force underlying SRCs, composite nuclear effects at quark resolutions, gluonic effects, sea quark effects, and exotic 6-quark states. Additionally, this new tensor polarized target opens the door to as-yet unmeasured tensor observables, such as the remaining three tensor structure functions $b_2,~b_3,$ and $b_4$, tensor DVCS that can access a new helicity term and further constrain sum rules, and tensor TMD functions that can potentially directly access a T-odd function~\cite{Bacchetta:2002xd}.
    Furthermore, the double helicity flip tensor structure function $\Delta(x,Q^2)$ allows to probe for novel gluonic components that are not present in nucleons.  A non-zero value would be a clear signature of exotic gluon states in the nucleus. 
    
    We close this section by mentioning the exciting possibility of utilizing a \emph{positron} beam on the polarized solid target.  The expected positron intensity in the proposed injector decreases from 5 $\mu$A at 10 MeV down to 100 nA at 60 MeV.  However the positron polarization increases from 10\% up to 75\% over this same energy range, thus opening up a sweet spot for use of the solid polarized targets with polarized positrons. 

\subsubsection{Polarized 3-He targets}


    The JLab polarized $^3$He target outlined in Section \ref{sect:poltarg}
    has undergone considerable development since its start in the 6\,GeV
    program and continues to break new ground in the 12\,GeV era.  This target
    allows unprecedented access to nuclear quark-gluon substructure by providing
    a highly polarized ($>55$\%) effective \emph{neutron} target capable of
    withstanding $e$-nucleon luminosities in the $10^{36}$--$10^{37}$ range
    with near-term designs.

    Recent measurements of $A_1$, $g_2$, $d_2$
    on the neutron using this target were undertaken in 2020--2021\cite{E12-06-110, E12-06-121}
    The combination of high luminosity and precision
    spectrometers allows us to push into the high-$x$, high-$Q^2$ region where
    model calculations for those observables are particularly sensitive to
    details of quark-gluon and quark-quark correlations within the nucleon.
    Constraints imposed by the current accelerator prevented a full exploration
    of those quantities, and there is interest in resuming and extending such measurements,
    particularly with the kinematic advantages of a higher energy electron
    beam.

    Measurements of $G_E^n$, and transverse single-spin asymmetries (SSAs)
    allowing extractions of Sivers and Collins functions on the neutron are on
    the horizon and are expected to provide insight into the nature of quark
    orbital angular momentum and flavor decomposition of nuclear PDFs.
    Hall C is the best, and perhaps only, facility capable of supporting this
    combination of high current polarized beam, a high luminosity polarized
    (effective) neutron target, with suitably matched high-precision detector
    systems.

    Future measurements exploiting a high-luminosity polarized neutron target
    may also include Deep Exclusive Meson Production (DEMP)  which provides insight
    into nuclear spin-flip GPDs (see also Sect.\,\ref{sect:precLT}), as well as 
    double-spin asymmetry (DSA) measurements which are sensitive to transverse momentum
    dependent (TMD) parton distributions.

For example, the transverse nucleon single-spin asymmetry in deep exclusive pion electroproduction, could be measured using longitudinal photons~\cite{PR12-12-005}. This observable is particularly sensitive to the nucleon helicity-flip pseudoscalar GPD, $\tilde{E}$, which, at small impact parameters can be interpreted in terms of correlated $q\bar{q}$ pairs in the nucleon's light-cone wave function.  Thus, this measurement is of fundamental interest to the QCD structure of the nucleon and a high scientific priority.  Since the measurement employs an L/T-separation (to isolate $\sigma_L$) in addition to the transversely polarized target, Hall~C remains the only feasible facility for such a measurement.  However, the development of a very high luminosity polarized $^3$He target would be required, such as a continuous flow $^3$He gas target based on a large volume polarizer and compressor, capable of at least 60\% polarization at a luminosity of $5 \times 10^{37}~\textrm{cm}^{-2}\textrm{s}^{-1}$ (80~$\mu$A beam current on a 10~cm cell)~\cite{PR12-12-005,Anderson:2020hlt}.  Such a target would open up many exciting possibilities for a new generation of hadronic structure measurements.


\subsection{A Definitive Measurement of the Real Part of the 2 Photon Amplitude}





\newcommand{\gevsq}{GeV$^2$}
\newcommand{\gevcsq}{(GeV/$c$)$^2$}
\newcommand{\qsq}{$Q^2$}
\newcommand{\GEp} {$G_E^p$}
\newcommand{\GMp}{$G_M^p$}

The discovery that the proton electric-to-magnetic form factor ratio decreases vs $Q^2$~\cite{MJ2000} suggests that the one-photon approximation for elastic electron-proton scattering may be inadequate for some observables and kinematic regimes. 
Many experiments and calculations have been performed on  \GEp/\GMp~over the last twenty years.
A recent paper~\cite{christy2021} provided the most up-to-date investigation of that ratio obtained via the Rosenbluth method.
Currently the discrepancy between the polarization transfer method and the Rosenbluth method is established on the level of 6-8 standard deviations.
At the same time, the full calculations of the \GEp/\GMp~ correction due to two-photon exchange (TPE) are incomplete due to the complexity of the hadron.

This situation motivated several measurements of the cross section ratio for electron/positron elastic scatting from a proton
for which the TPE effect calculation is supposed to be simpler.
Electron/positron type searches for the TPE effect were performed at SLAC and other labs about 50 years ago but
were inconclusive due to insufficient precision~\cite{SLAC}.
Several years ago, three new experiments were completed and the TPE in elastic scattering was observed~\cite{IR2015, BH2017, DR2017}.
However, the precision of those measurements is a bit low and, due to relatively low \qsq~of 1-2 \gevsq, the theoretical predictions are not stable.
Currently, a new proposal for a TPE experiment is under development which would employ the pulsed positron and electron beams at DESY~\cite{OL2019}.

\begin{figure*}
\centering
\includegraphics[width=0.75\textwidth]{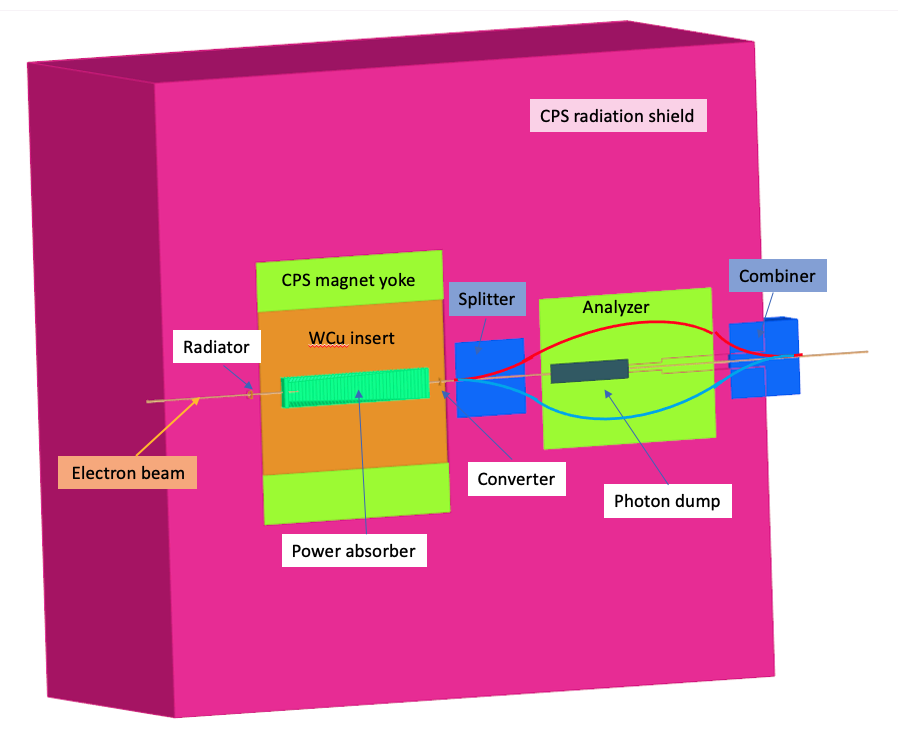}
\caption{The layout of the positron/electron beam source with CPS.}
\label{fig:source}
\end{figure*}

The most difficult challenge of such experiments is controlling systematics in the difference between the electron and positron beams.
Most experiments of this type have alternated between a beam of positrons and a beam of electrons every few hours.
The experiment in CLAS6~\cite{DR2017} used a mixed beam, which in principle allows parallel data taking with both beams.
However, the toroidal magnetic field in the detector system of CLAS6 created an "in-bending vs out-bending" asymmetry between the scattered leptons hence new systematic issues. 

 
We will propose a mixed beam experiment in Hall C using a very  different configuration of the detector system. 
The system will use a high resolution calorimeter for the recoiling proton and a magnetic spectrometer
for accurate measurement of the direction and momentum of the scattered positrons and electrons.
The mixed beam will be derived from interaction of the CPS photon beam with a $\gamma \rightarrow e^+e^-$ converter. 
The convertor will be located 1 meter downstream from the CPS radiator to ensure a small (1 mm) diameter for the $e^+e^-$ beam 
at the production point. 
Following creation of the $e^+e^-$ pairs, they will pass through a chicane to select a high momentum bite, and the remainder of the photon beam will be dumped. 
This compact configuration will allow a short distance and therefore a relatively small lepton beam spot on the physics target.
The concept for the broadband mixed $e^+e^-$ source is illustrated in Fig.~\ref{fig:source}.

The key parameter of the beam source is its intensity, which can be estimated directly using the electron beam power 
and the thicknesses of the radiator and convertor. \\
\begin{equation}
I_\pm = \frac{P}{E_\textrm{beam}} \times t_1 \cdot \frac{\Delta E_\gamma}{E_\gamma} \times t_2 \cdot \frac{\Delta E_\pm}{E_\pm},
\end{equation}
where $I_\pm$ is the intensity of the positron/electron, $P$ is the power of the primary electron beam, 
$\Delta E_\gamma$ is the full range of photon energies from $E_\textrm{beam}$ down to $E_\pm$ which can produce leptons with given energy, 
$E_\textrm{beam}$ is the energy of electrons in the primary beam, $t_1$ is the radiator thickness, $t_2$ is the convertor thickness, and 
$\frac{\Delta E_\pm}{E_\pm}$ is the relative acceptance for beam energy of the secondary particle. 
In our case the P is 30 kW, $t_1=t_2 =0.1$ radiation length, $\frac{\Delta E_\gamma}{E_\gamma}$ is about 2.6/6.6 = 0.4, and
$\frac{\Delta E_\pm}{E_\pm}$ is 0.2.
For the experiment with incident 4 GeV leptons we propose to use a primary electron beam with 6.6 GeV.
The value of $\frac{\Delta E_\pm}{E_\pm}$ was selected based on the projected acceptance of the detectors.
The resulting positron flux is $2 \times 10^{10}$ per second (about 3nA), roughly a factor of 250 higher than was used 
in the CLAS6 experiment. 
However, for CLAS6 a much larger $\frac{\Delta E_\pm}{E_\pm}$ could be used, so the effective advantage factor for this proposed Hall C configuration will be about 15.

We propose to do measurements at \qsq=3~\gevsq~at several kinematic points.
Using $\frac{\Delta Q^2}{Q^2} = 0.1$, we found the projected counting rate using BigBite to be 2.5 Hz at a scattering angle of 32$^\circ$ and $\epsilon = 0.77$.
With 10 hours of production time, e$^\pm$p event statistics will be 90k events.
Lower lepton energy is needed for a data point at a smaller epsilon value.
For 2.5 GeV central lepton energy and a 70$^\circ$ scattering angle, the data will have $\epsilon = 0.36$ 
and the projected rate is about 0.5 Hz.
This will require 50 hours for 90K statistics.
The statistics need to be collected for a few polarity flips in the beamline magnets and lepton spectrometer.
Overall beam time for a sub 0.5\% accuracy experiment is about 500 hours.

In this section we have sketched how to make a definitive measurement of the real part of the two-photon amplitude in the elastic scattering of electrons and positrons from the proton. 
The very high power photon beam requires the CPS, and detection of the scattered particles requires medium acceptance devices such as BB and SBS. Using a deuterium target, such measurements could be extended to the neutron by employing HCAL to detect quasi-elastically scattered neutrons (as well as protons as a control). Currently, Hall C is the only facility where such a program could be realized. 





 
%

\label{CPS_e+}
 
\subsection{Recoil Tagging Program}

\subsubsection{Sullivan process}
\begin{figure}[hbt!]
  \centering
\includegraphics[width=0.5\textwidth]{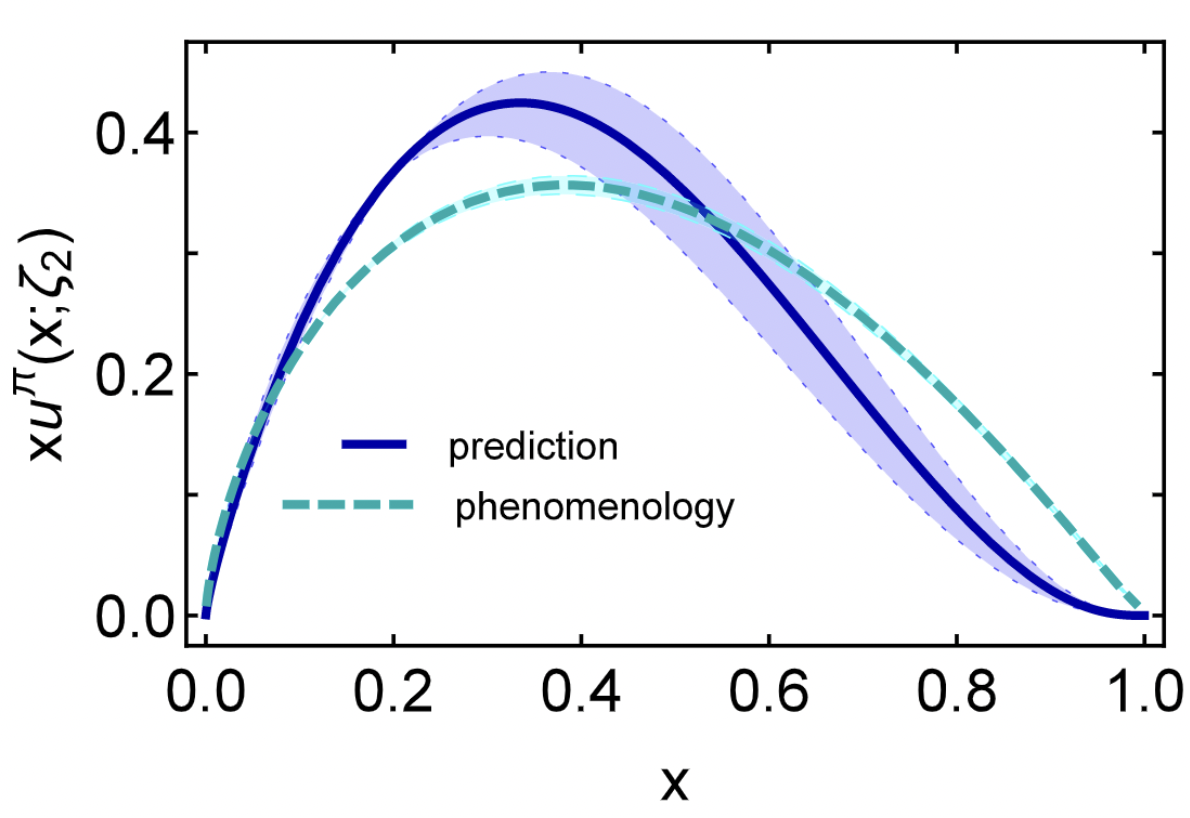}
 \caption[]{Reproduced from Ref.~\cite{roberts21}, the solid blue lines shows the calculated pion valence-quark distribution with $\zeta_2$~=~2~GeV as the scale at which the pion is being resolved. The blue band is the estimated uncertainty of the calculation. The dashed line shows a fit obtained by analysing data on $\pi$-nucleus Drell-Yan and leading neutron electroproduction data~\cite{Barry18}.}
  \label{fig:pionsf}
 \end{figure}

Novel experiments to precisely determine the mesonic content of the nucleon, as well as pion and kaon structure, are planned for Hall C and the EIC in the near term. Such information is accessed by tagged scattering from the “meson cloud” of the nucleon via the Sullivan process. Paradoxically, these lightest pseudoscalar mesons
appear to be the key to the further understanding of the emergent mass and structure of hadrons. They are the Nambu-Goldstone boson modes of QCD, associated with dynamical chiral symmetry breaking. The interference between the emergent hadron mass and the Higgs-boson mass mechanisms is seen to be responsible for $\sim$95\% of the pion's mass and $\sim$80\% of the kaon's mass~\cite{roberts21}. Unraveling the partonic structure of mesons, and the interplay between emergent and Higgs-boson mass mechanisms, is a common goal of three interdependent theoretical approaches: continuum QCD phenomenology, lattice-regularized QCD, and the global analysis of parton distributions – all of which are linked to experimental measurements of hadron structure~\cite{Cosyn:2016}. For all, and for global analyses in particular, mapping tagged
pion and kaon cross sections over a wide range of kinematics will be critical. Moreover, theoretical calculations of the pion structure in the valence region tend to disagree with each other. (See Fig.~\ref{fig:pionsf} for one such calculation compared to a phenomenological fit to existing data.) This tells us that it is essential to measure the pion structure function over a wide range of $x$ using new techniques. The pioneering 12 GeV TDIS measurements in Hall A at low $Q$ and large $x$ will complement the planned high $Q$, low $x$ EIC studies and the scant existing HERA data. 

Such measurements can also be carried out in Hall C using the compact solenoidal spectrometer discussed in Sec.~\ref{sec:css}, or one of the existing spectrometers. Tagging the meson by detecting the spectator nucleon at low momentum may be achieved over a wide range in angle using specialized detectors placed around the target, such as the multi-TPC based system planned for Hall A. These experiments would measure semi-inclusive reactions $H(e,e^{\prime}p)X$ and $D(e,e^{\prime}p p)X$ in the deep inelastic regime over a wide range of $Q^2$ and $x$, for very low proton momenta. The key to this experimental technique is to measure the low-energy outgoing recoil proton in coincidence with a deep inelastically scattered electron from the hydrogen target. In the deuterium case, an {\it additional} low energy spectator proton will be identified at backward angle to tag the neutron target. The inclusive electron kinematics determine that a DIS event has occurred, and the low momentum protons measured by the recoil detector, in time and vertex coincidence with the DIS event, ensure that the deep inelastic scattering occurred from partons within the meson cloud of the nucleon. The contributions of the meson cloud to the structure function of the nucleon will be studied in terms of the ratio of semi-inclusive structure function $F_2^{(\pi p)}$ to the inclusive nucleon structure function $F_2^p$ distributed in $x$, and recoil proton momentum and angle. The pion structure function can be extracted from these experiments using phenomenological models of the meson cloud~\cite{Hobbs,Holtmann96,Melnitchouk:1995en} and comparing to pionic Drell-Yan data where they overlap in $x_{\pi}$.

The small cross sections involved in these measurements dictate a need for high luminosity experiments. The initial planned experiment in Hall A~\cite{TDIS} will be both $Q^2$ limited and pioneering. Continued studies at higher $Q^2$ in Hall C, enabled by higher electron beam energies, will begin to fill out kinematics and bridge the gap to the lower $x$ EIC data as well as any eventual real pion measurements from COMPASS. 
These unique tagged DIS experiments will help measure the mesonic content of nucleons and the pion structure function. The results of these experiments will establish a precision frontier and  confirm a long standing prediction of nuclear physics.
 
\subsubsection{Nuclear tagging}
Extending the tagged DIS measurements to light nuclei such as $^3$He and $^4$He will enable the study of nuclear modification of the semi-inclusive structure function $F_2^{(\pi p)}$. Such measurements will use the multi-TPC recoil detector to explore the nuclear modification of the mesonic content of nuclei and, along with the compact solenoid spectrometer discussed in Sec.~\ref{sec:css}, they can cover a wide range of kinematics. These nuclear tagging experiments have the potential for uncovering the long sought after "nuclear pions". Furthermore, with the extraction of the pion structure function in light nuclei, they will pioneer the first EMC type measurement for mesons.

The space-time evolution of the strong interaction can be probed uniquely by examining FSI in DIS off nuclei. Typically, such studies measure the attenuation of leading hadrons produced in the current fragmentation region~\cite{Adams94,Ashman91,Airapetian01}.  An alternative is to study the production of low-energy hadrons emitted from the target remnant (i.e. in the nuclear fragmentation region)~\cite{Larionov20}. Soft neutron production with energies $E <$ 10~MeV in DIS off nuclei is one of the very few observables which can be studied both at fixed target experiments and at an $eA$ collider. Therefore, the study of soft neutron production would provide a new, sensitive probe of the dynamics of FSI’s in DIS which could prove to be a benchmark for comparison and interpretation of results between JLab and the EIC. Extending the spectator tagging measurements to light nuclei with the help of high resolution multi-TPC recoil detectors for soft neutron tagging, would facilitate this critical synergy between the two labs. The high luminosity possible in Hall C is essential given the small cross sections of these measurements. Moreover, the large range of $Q^2$ and $x$ enabled by higher electron beam energies will be complementary to the lower $x$ EIC data.

\subsection{Constraining strange quark elastic form factors at high momentum transfer}\label{sec:COP}
 





\newcommand{\tc}{,~}
\newcommand{\GE} {$ {G_{_{_E}}}$}
\newcommand{\GD}{$G_{\textrm{Dipole}}$}
\newcommand{\gn}{\mbox{$\mu_n{ G_{_{\!E}}^{n}}$}/\mbox{${ G_{_{\!M}}^{n}}$}}
\newcommand{\Fone}{${ F_1}$}
\newcommand{\Ftwo}{${ F_2}$}
\newcommand{\Fonep}{\mbox{${ {F_1^{\it p}}}$}}
\newcommand{\Ftwop}{\mbox{${ {F_2^{\it p}}}$}}
\newcommand{\Fonen}{\mbox{${ {F_1^{\it n}}}$}}
\newcommand{\Ftwon}{\mbox{${ {F_2^{\it n}}}$}}
\newcommand{\Foneu}{\mbox{${ {F_1^{\it u}}}$}}
\newcommand{\Ftwou}{\mbox{${ {F_2^{\it u}}}$}}
\newcommand{\Foned}{\mbox{${ {F_1^{\it d}}}$}}
\newcommand{\Ftwod}{\mbox{${ {F_2^{\it d}}}$}}
\newcommand{\Fsp}{\mbox{${ {F_{\it s}^{\it p}}}$}}
\newcommand{\Sp}{${ {Q^2 F_2^{\it p}/F_1^{\it p}}}$}
\newcommand{\Sn}{${ {Q^2 F_2^{\it n}/F_1^{\it n}}}$}

\newcommand{\PR}{{ Phys. Rev. }}
\newcommand{\PRL}{{ Phys. Rev. Lett. }}
\newcommand{\PL}{{ Phys. Lett. }}
\newcommand{\NP}{{ Nucl. Phys. }}
\newcommand{\NIM}{{ Nucl. Instr. Meth. }}
\newcommand{\etal}{{\em et al.}}
%
%
%



\begin{figure}[ht]
      \includegraphics[trim = 0mm 0mm 0mm 0mm, width=0.65\textwidth] {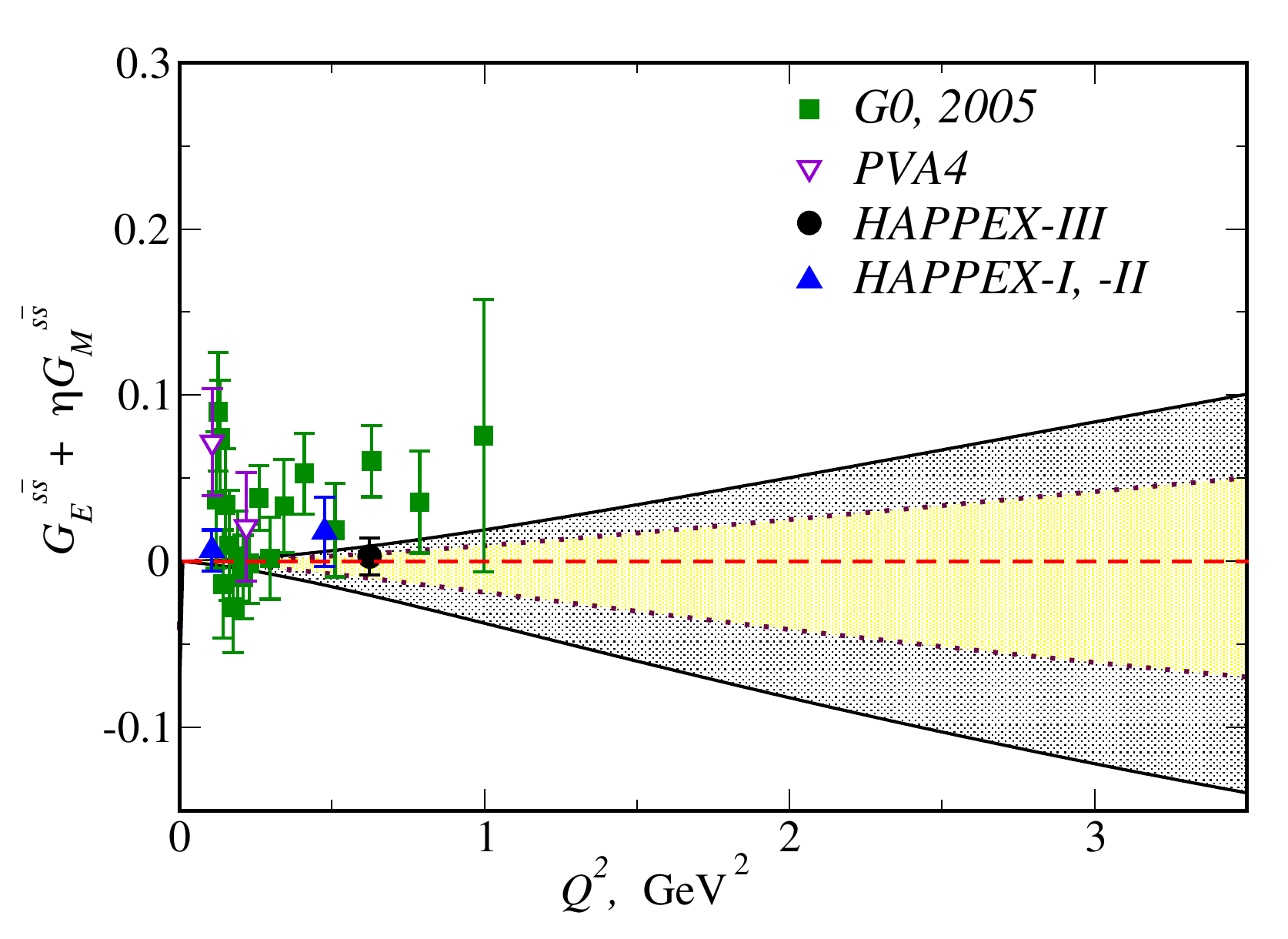} 
     \caption{ The strange form factor vs. $Q^2$ data and projections per Refs.~\cite{TH2015,TH2019}. The yellow band represents a 1 sigma limit while the gray band corresponds approximately to a 4 sigma limit.}
     \label{fig:TH}
\end{figure}

The strange quark content of a nucleon wave function became a prominent subject with the discovery of the EMC effect and follow-up theoretical investigations~\cite{KM-1988}.\footnote{This section is an abbreviated version of material presented at the ECT* 2019 workshop, https://arxiv.org/abs/2001.02190 . }  
The method for extracting strange quark form factors from a combination of parity conserving and parity violating elastic scattering measurements was formulated in Refs.~\cite{BM-1989, DB-1989}.
Studies were performed by the SAMPLE and G0 experiments and HAPPEXs and PVA4. (See data points in Fig.~\ref{fig:TH}.) 
High accuracy results were obtained for \qsq~less than 0.6 GeV$^2$, and much less precise results for \qsq~up to almost 1~\gevsq.
However, the role of \Fsp~could be larger at several \gevsq.
For example, the neutron electric form factor is zero 
at \qsq=0 yet grows relative to the dipole form factor up to 2-3 GeV$^2$. 
A recent analysis of the possible value for the strange form factor
suggests that \Fsp~could be even larger than \GD, or 0.03 at $Q^2$=3.4~\gevsq. (See Fig.~\ref{fig:TH}.) 

In a single measurement, the parity non-conserving asymmetry has contributions from the 
electrical and magnetic form factors from strange quarks and axial nucleon form factors.
Quasi-elastic scattering of the electrons from a deuteron allows us to constrain the axial form factor as discussed in Ref.~\cite{BB2005}.
Here we present specific ideas for a new experiment to determine {\Fsp} in a coincidence measurement using SBS equipment.
The proposed measurements on the proton and deuteron targets will provide empirical constraints on those form factors in a higher \qsq~range which was not explored before.


 
%


The nucleon form factors for the virtual photon have three contributions:
\begin{equation}
G^{\gamma}_{\rm^{E,M}} = {\textstyle{2\over3}}G^u_{_{E,M}} 
- {\textstyle{{1\,}\over{3}}} G^d_{_{E,M}} 
- {\textstyle{{1\,}\over{3}}} G^s_{_{E,M}} 
\label{eq:GEM}
\end{equation}

\noindent The Z boson analogs of Equation \ref{eq:GEM} have a similar structure, but with the quark charges replaced by the vector weak charges of the quarks.
A new measurement of the parity violating asymmetry of longitudinally polarized electrons scattering from a proton at high $Q^2$,  combined with the existing world data for $G^{\gamma}_{\rm^{E,M}}$ for the proton and neutron, the assumption of isospin symmetry, and constraints on the axial contributions from other experiments, will enable a complete flavor decomposition of the nucleon form factors and determine a linear combination of $G^s_{_{E,M}}$~\cite{DB2001}. 
 

%


There are two experimental difficulties in doing the \Fsp~measurement at large \qsq: 
the relatively low counting rate and the large background from inelastic electron-proton scattering.
The reduction of the counting rate, which is due to reduction of the $\sigma_{_{\textrm{Mott}}}G^2_{_{\textrm{Dipole}}}$, 
is partly compensated by a linear increase of the total asymmetry for high \qsq.
To suppress the inelastic events, one can use the tight time and the angular correlations between the scattered electron and recoil proton as was proposed in Ref.~\cite{BW2005}.

\begin{figure}[ht]
\unitlength 1cm
\begin{minipage}[th]{10cm}
   \includegraphics[width=1.0\textwidth] {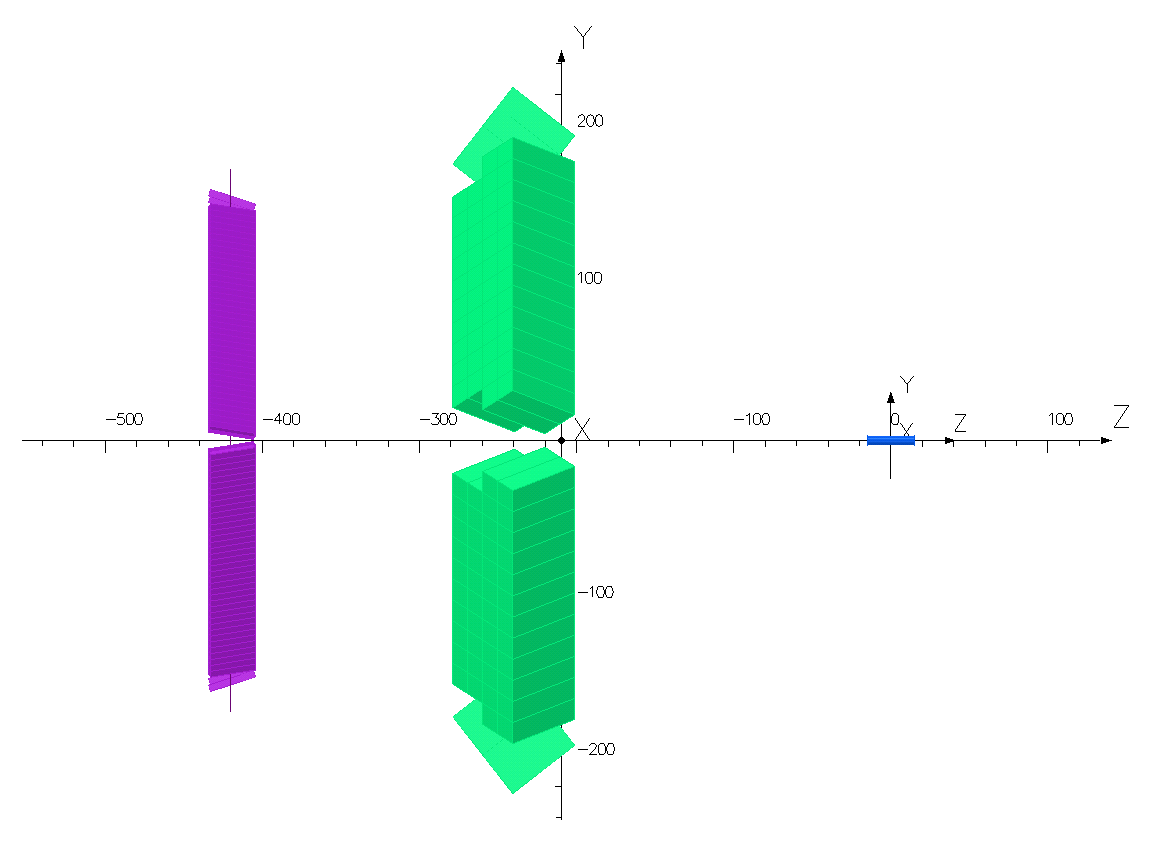} 
\end{minipage}
\begin{minipage}[th]{10cm}
\includegraphics[width=0.9\textwidth] {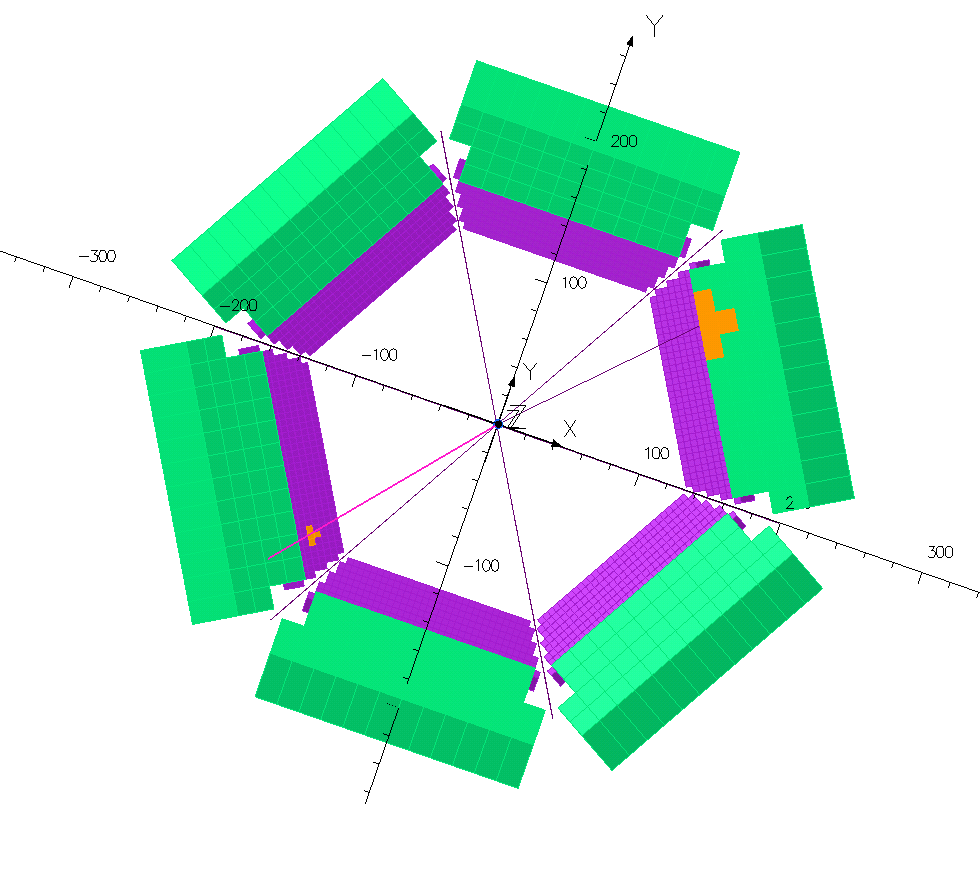}
\end{minipage}
   \caption{ Top panel: Side view of the apparatus. The electron beam goes from right to left.
   The proton detector is shown in green and the electron detector in purple; the liquid hydrogen
   target is shown in blue. Bottom panel: Front view of the apparatus. The blocks in orange get signals
   from the electron and the proton whose directions are shown in red.} 
   \label{fig:COP2}
\end{figure}

The solid angle of the apparatus should cover a suitable range of the momentum transfer $\Delta Q^2/Q^2 \sim 0.2$ for which the event rate variation over the detector solid angle is acceptable.
The equipment needed for such an experiment could be obtained from the SBS where a highly segmented
hadron calorimeter and electromagnetic calorimeters will be used for the GEp experiment~\cite{GEpE}.
Fig.~\ref{fig:COP2} shows the proposed configuration of the detectors.
Modern electronics such as a flash ADC allows dead-time free data collection.
 
The proposed detector configuration has an electron arm with a solid angle of 0.06~sr at a scattering angle of 15.5$\pm$1.5 degrees.
Within 30 days of data taking with a 6.6 GeV beam, the PV asymmetry will be measured to 3\% relative accuracy which corresponds to an uncertainty of \Fsp~of 0.002.
(Work is in progress to determine to what extent the uncertainty of the long distance scale corrections to the axial asymmetry will limit the interpretability.)  
Such a measurement will provide the first experimental limit on \Fsp~at the large momentum transfer 
of 2.5~\gevsq~(or discover its non-zero value) and reduce the current uncertainty from the strangeness contribution 
in the flavor separated proton form factors such as \Ftwod~by a factor of 6. 

The concept above is close to the one in Ref.~\cite{BW2005} but relies on the existing hadron calorimeter, existing radiation hard electron calorimeter, and advanced DAQ currently available.
The JLab PAC29 had a concern about the impact of the proton transverse polarization, see~\cite{PAC29}, on determination of the proton coordinate in the shower detector.
A recent study~\cite{BLITSTEIN} found that polarization of the recoil proton will shift the detected coordinate by $\sim$10~$\mu$m, in close agreement with our analytical estimate. 
Such a value of the coordinate shift contributes 1\% experimental uncertainty in PV asymmetry.

In the planned kinematics, the PV asymmetry is O(100) ppm. Achieving a few percent statistical uncertainty to constrain the strange quark form factors therefore demands the experiment acquire $10^{11}$ events at high momentum transfer. A high luminosity facility is absolutely required. Due to constraints from the MOLLER and SOLID programs in Hall A, it is likely that Hall C will be the only available hall for a large installation experiment such as this.

%

\subsection{Precision Measurement of Hyperon Decay \label{hyperondecay}}
Almost five decades after the discovery of $\Lambda^0$ polarization in inclusive unpolarized p-Be scattering~\cite{Bunce76}, the exact physical origin of this spontaneous (self) polarization phenomena remains a mystery. Quark helicity conservation and the small magnitude of spin-flip amplitudes arising from higher twist contributions, along with the high multiplicity of inclusive reactions, imply that the large self polarizations cannot be understood in perturbative QCD. This is just as serious a problem as the nucleon spin problem, and the two issues are probably closely related~\cite{Troshin02}. It even led to the introduction of new type of leading twist fragmentation functions (FF) called polarizing FF that depend on the quark transverse polarization~\cite{Anselmino}. The hyperon polarization is typically determined from its  self-analyzing parity violating weak decay by measuring the angular distribution of the decay products. These self-analyzing hyperon decays are a strangeness key to study confinement in QCD. This is primarily because the mass of the strange quark ($\sim$ 95 MeV/c$^2$) is closest to the QCD cut-off $\Lambda_{\textrm{QCD}} \approx $ 200 MeV/c$^2$, the scale where quarks are confined into hadrons.

Triple coincidence measurements with the HMS, SHMS and NPS can provide a unique precision window into hyperon self polarization. In the case of the $\Lambda^0$ and $\Sigma^0$ decay, the seldom used $\Lambda^0 \rightarrow   \pi^0 n$ and $\Sigma^0 \rightarrow \Lambda^0 \gamma$ decay channels can be employed. The triple coincidence reactions p($\vec{e}, e^{\prime}K^{+}\pi^0$)n  and p($\vec{e}, e^{\prime}K^{+}\gamma$)$\Lambda^0$ can be measured with the scattered electron detected in the HMS and the $K^{+}$+$\pi^0/\gamma$ detected in the SHMS+NPS combined spectrometer system. The polarized electron beam and the parity violating weak decay of the hyperons imply that in the rest frame of the hyperon the angular distribution of the decay products is given by~\cite{PDG2020};
$$ \frac{dN^{\pm}}{d\cos{\theta}_{\pi^0/\gamma}} = N\left[ 1 + \alpha(P^0 \pm P_bP^{\prime})\cos{\theta}_{\pi^0/\gamma} \right], $$
where $\theta$ is the C.M. angle of the decaying $\pi^0/\gamma$, $\alpha$ is the parity violating decay parameter, $P^0$ is the induced self polarization, $P_b$ is the beam polarization, and $P^{\prime}$ is the polarization transfer. The high luminosity and the high resolution spectrometers of Hall C will enable precision measurement of $P^0$ and $P^{\prime}$ in hyperon decay over a large range of $W, t$ and $Q^2$. This will allow precision tests of the strange quark spin, orbital angular momentum and final state interactions and provide unique insight into confinement in QCD. Similar measurements with nuclear targets will provide precision information on the hyperon response functions $R_{TL'}$ and $R_{TT'}$ and their $A$ dependence. All of this will help unravel the five decade old puzzle of hyperon self-polarization.

In addition, the measurement of a non-zero angular asymmetry in the $\vec{\Sigma}^0 \rightarrow \Lambda^0 \gamma$ decay via the reaction p($\vec{e}, e^{\prime}K^{+}\gamma$)$\Lambda^0$ 
would serve as an indication of strong CP violation beyond the SM~\cite{Nair19}. This reaction provides an alternate test of CP violation that is a hybrid of the traditional weak decay of baryons and electric dipole moment (EDM) searches. It was recently pointed out that this angular asymmetry has never been measured~\cite{Nair19}. Given the challenges of measuring $\Sigma^0$ decays, a high luminosity facility such as Hall C would be very helpful. 

Furthermore, with a focal plane polarimeter (FPP) in the SHMS, one can measure the parity violating decay parameter of the $\Sigma^{+}$ hyperon with unprecedented precision. Note that the PDG value of this decay parameter is based on a measurement from the 1960s with less than 1500 events~\cite{PDG2020}. The triple coincidence reactions p($\vec{e}, e^{'}\vec{p}\pi^0$)K$^{0}$ can be used to measure the decay parameters of the $\Sigma^{+}$ hyperon ($\vec{\Sigma}^{+} \rightarrow p \pi^{0}$). The polarization transfer to the recoil proton has a well defined dependence on ${\bf P}_{\Sigma}$ (the polarization of the $\Sigma^{+}$ hyperon) and the three possible orientations with respect to the recoil protons C.M. momentum direction, $\hat{n}$, and the corresponding decay parameters, $\alpha$, $\beta$, and $\gamma$~\cite{donoghue86,PDG2020}.
The angular distribution of the recoil proton is related to the $\Sigma^{+}$ polarization as~\cite{donoghue86};
$$\frac{dN^{\pm}}{d\cos{\theta}_{p}} = K(1 + {\bf P}_{\Sigma}\cdot \hat{n}),$$
and can be measured with high precision with the SHMS. The parity violating weak decay parameters $\alpha$ and $\beta$ can then be extracted using the well established technique of measuring the two components of the recoil proton polarization (${\bf P}_p$) in the 
FPP. 
The high precision measurement of these critical SM parameters is interesting in their own right, in addition the direct measurement of the ratio of the decay parameters $\beta/\alpha$ is  a sensitive test of CP violation~\cite{donoghue86}. Thus, a program to measure hyperon decays utilizing the high luminosity capabilities of Hall C can provide critical insight into the confinement problem and also serve as a probe of physics beyond the SM.

\section{Potential Significant Accelerator Upgrades}

Significant upgrades of the accelerator would allow the Hall C program to expand into further exciting new areas. Two that have been under discussion are the possibility of providing the halls with positron beams or electron beam energies as high as 20 GeV.

\subsection{Positron Beam Facility at 12 GeV}
A year-long conceptual design study of a continuous-wave polarized
positron source at CEBAF is underway.  A new approach for generating
polarized positrons would be exploited at Jefferson Lab: referred to as
PEPPo (Polarized Electrons for Polarized Positrons), the approximately 90\%
polarization of the CEBAF electron beam is transferred by the
electromagnetic bremsstrahlung shower within a high-Z target to e+/e-
via pair creation.    The baseline design for $e^+$ @ CEBAF 12 GeV begins
with a 50-100 MeV and 2 mA
electron beam
to generate the positron shower.
A useful “slice” of the transverse-longitudinal 6D distribution is
selected using a combination of transverse focusing magnets, charge
separating dipoles, acceptance limiting collimators, and RF cavities
for bunching and acceleration.

Positrons bunches are then re-accelerated to 123 MeV and injected into
CEBAF for acceleration to 11 GeV, where magnets are operated with
reversed polarities.  Despite a much larger initial momentum spread,
positron beams benefit from adiabatic damping which
results in the same momentum spread as electron beams.  The  large
positron  beam  emittance  at  the  injector  entrance  is  also
strongly reduced by acceleration effects which result in a final beam
size at worst 2-3 times larger than the electron beam from the standard source.  
Until a more developed proposal of the positron injector is agreed upon, proposals have been directed by the Jefferson Lab Positron Working Group to consider as a base-line the availability of unpolarized beams up to $3~\mu\textrm{A}$ and polarized beams up to 100 nA with >50\% polarization.  
To control systematic errors in these experiments, the positron source should be configured to allow selection of either the positron or the electron from pair production.
This way the beams have the same emittance and encounter the same forces throughout the accelerator. 


  
\subsection{Beam Energy Upgrade to 20 GeV}
A working group is examining the feasibility of increasing CEBAF energy to 20 GeV.  This would be accomplished by increasing the injection energy to 650 MeV and replacing the last electromagnetic pass with a permanent magnet FFA (Fixed Field alternating gradient Accelerator) which would contain 5.5 passes.  The existing first pass electromagnets would be eliminated and the four higher energy passes raised to allow the higher injection energy using existing magnets. 

The ratio of the injector energy to last pass energy into the North Linac is critical to maintaining tolerable beam sizes in the North Linac. For the 12 GeV machine, this ratio is 10975/123 = 89. The last two passes through the North Linac are effectively drifts. This must change for the further increase of energy.  With 20 GeV entering the North Linac on the last pass (to Hall D), the energy ratio will be 31:1 and the beam envelope can be adequately constrained.  The energy ratio of the multiple beams in the South Linac will increase from 8 to 9.

For reasons of cost, and to minimize the impact on operation of the 12 GeV CEBAF, it has been decided that first the positron injector and later the 650 MeV energy upgrade injector will be housed in the building which once housed the JLab free electron laser (FEL).  A kilometer transfer line will direct first the positrons and then the 650 MeV electrons to the front of the North Linac where a new vertical merger of the injection beam with the recirculated beams will take place.  The beginning of this transfer line will be housed in a new tunnel from the FEL building to the South Linac stub.  The majority of the line will be hung from the ceiling of the existing tunnel. 

A preliminary design of a conventional electromagnetic dipole for the beamlines leading to the halls has been completed. Extraction region real estate is quite restricted and unconventional devices, including conductively cooled superconducting RF separators and septum magnets, will be needed. Detailed optics work is underway.

With the beam energy upgrade, experiments studying exclusive reactions will need to reach higher momentum and smaller angles than can presently be accessed by the HMS. 
The HMS can be upgraded to 
accept particles with momentum of  $20~\textrm{GeV}/c$ and reach small scattering angles.
The present HMS was built conservatively to reach $7.2~\textrm{GeV}/c$ by using 2 Tesla magnets with long effective field lengths 
and plenty of drift space. To reach 20~GeV/c, the HMS magnets would need to operate at 6 Tesla which 
can be achieved with present day superconducting magnets. As with the SHMS, a horizontal bending magnet would be needed to allow the HMS to access small scattering angles. Upgrading HMS would involve removing the present magnets and overhead steel structure, procuring a new suite of magnets, and installation. The entire HMS pivot assembly, support structure and shield house could be reused. For the experiments presented in this paper, the present HMS momentum range is sufficient. 
We encourage physics proposals that would need an expanded HMS momentum range to help define specifications for such an HMS upgrade.

 



\section{Summary }

Hall C has played an important role in the understanding of hadron structure by offering a unique facility optimized for precision measurements of small cross sections and a flexible configuration with many opportunities for new experimental equipment. Notable results to date include increased understanding of the quark structure of nucleons and nuclei through precision measurements of inclusive and semi-inclusive cross sections and precision measurements of the pion form factor.  Over the next decade, a vibrant physics program is envisioned addressing physics topics that may be broadly categorized as the spatial distribution of quarks inside nucleons and nuclei, the quark structure of nucleons and nuclei in momentum space, the origin of nucleon spin, the nature of confinement/hadronization, the origin of hadron mass, hard/soft QCD factorization, and the nature of the strong/nuclear force. Hall C Future Science can be summarized as ``high precision and versatility at the luminosity frontier''. The potential upgrades to the accelerator of a positron beam and higher beam energy will offer  unique opportunities for accessing new physics topics and further increase the kinematic reach of existing physics topics.
 

%
  
Further detailed information about Hall C and its User Group can be found online at the links \href{ https://www.jlab.org/physics/hall-c }{Hall C home} and \href{ https://hallcweb.jlab.org/wiki/index.php/User_Group }{Hall C User Group}, respectively.  
The flexible way in which Hall C experiments are reconfigured from year to year provides a fantastic learning opportunity for students and postdocs.


\section*{Acknowledgements}
This work was supported in part by Department of Energy (DOE) contract number DE-AC05-06OR23177, 
under which the Jefferson Science Associates operates the Thomas Jefferson National Accelerator Facility. This work was support in part by NSF grants PHY-2012430, PHY-1530874,
and NSERC grant SAPIN-2021-00026. 
The HCAL project was accomplished by the collaboration of CMU, Catania University, and JLab with funding support from DOE and INFN.

\appendix

\section{Polarimetry Trade-offs at 20 GeV \label{AppPol20GeV} }

Operation of both Hall C polarimeters up to a possible 20~GeV poses significant
challenges.  The dipoles in the Hall C Compton polarimeter have sufficient field strength to operate slightly higher than 11 GeV, but operation at 20 GeV would require either further reducing the 13 cm
vertical deflection, or lengthening the chicane.  The former option would not allow the use of a photon detector since it is
already very challenging to make a Compton photon detector that is sufficiently compact that it can fit into the relatively small
vertical space available with a 13 cm deflection.  A system that only uses an electron detector is possible, however such a system
would pose challenges during initial beam setup (since photon detector rates are used to optimize the beam-laser relative position)
and would have limited applicability at lower beam energies.
Lengthening the Compton chicane would also pose challenges: assuming
the same deflection (13 cm), and dipole fields comparable to those required for operation at 11 GeV,
the dipole chicane length would have to be increased from 11~m to 17~m.  With no additional changes, this much space is not
available on the Hall C beamline.

The Hall C M\o ller polarimeter faces similar issues.  The existing magneto-optical system does not have sufficient
strength to operate at 20~GeV in the current configuration.  Reducing the horizontal detector position from 49~cm to 20~cm
and increasing the drift from the last quad to the detectors by about 4 meters would allow use of the existing magnets.
Detailed simulations would be required to determine the impact of the new detector positions on the backgrounds in
the system.

 
\bibliographystyle{apsrev4-2}
\bibliography{refs.bib}


 




\end{document}